# Continuous-Time Aggregation of Massive Flexible HVAC Loads Considering Uncertainty for Reserve Provision in Power System Dispatch

Jingguan Liu, *Student Member, IEEE,* Xiaomeng Ai, *Member, IEEE,* Jiakun Fang, *Senior Member, IEEE,* Shichang Cui, *Member, IEEE,* Shengshi Wang, *Student Member, IEEE,* Wei Yao, *Senior Member, IEEE,* and Jinyu Wen, *Member, IEEE*

***Abstract*—Heating, ventilation, and air conditioning (HVAC) loads, with their rapid response capabilities, can provide considerable intra-hour flexibility on the demand side for reserve provision in order to follow the fast variations of renewables. However, scheduling massive HVACs is challenging due to computation complexity and the uncertainty of outdoor temperature. In this paper, we first introduce a novel continuous-time (CT) aggregation model to reveal the potential intra-hour flexibility of HVACs. For accurate aggregation, a new affine transformation is designed to handle the heterogeneity in high-dimensional feasible region. Further, for reliable aggregation in practical environment, the outdoor temperature uncertainty is constructed by distributionally robust chance constrains and integrated into the aggregation model. Secondly, for the tractable calculation of the proposed CT aggregation model, a cascade of tailored reformulation techniques is proposed, including the Bernstein polynomial spline, polytope projection, and linearization transformation. Thirdly, a customized hierarchical dispatch framework is proposed via incorporating the proposed CT aggregation model into reserve provision in power system dispatch, so as to efficiently schedule massive HVACs to cope with the renewable uncertainty. Case studies verify the effectiveness and scalability of the proposed CT aggregation model in aggregation accuracy, intra-hour flexibility utilization, and uncertainty handling.**

## NOMENCLATURE

In this paper, main symbols and notations are clarified below for quick reference. Others will be defined when they first appear, if desired. Also, we use italics for single parameter or variable, bold fonts for matrices or vectors, and calligraphic fonts for sets.

*Abbreviations:*

| | |
|---|---|
| HVAC | Heating, ventilation, and air conditioning. |
| BP | Bernstein polynomial. |
| DRCC | Distributionally robust chance constraint. |
| DT/CT | Discrete-time/continuous-time. |
| M-sum | Minkowski sum. |

*Indices and Sets:*

| | |
|---|---|
| $\tau$ | Time instant. |
| $i/j$ | Bus index. |
| $k/N^K$ | HVAC index/number. |
| $s$ | Scenario index. |
| $t/N^T$ | Period index/number. |
| $\mathbb{U}^{agg}/\mathbb{U}^{app}$ | Exact/Approximate aggregated set of massive HVACs. |
| $\mathbb{U}^{base}/\mathbb{U}$ | Base/Exact set of single HVAC. |
| $\mathbb{U}^{aff}$ | Affine-transformed base set. |

*Parameters:*

| | |
|---|---|
| $a^{c1}/\dots/a^{c7}$ | Thermal coefficients for HVACs. |
| $T^o$ | Outdoor temperature. |
| $H^a/H^m$ | Thermal conductance of indoor air/mass. |
| $C^a/C^m$ | Thermal capacity of indoor air/mass. |
| $\mu^h/f^h$ | Coefficients for HVACs in [0,1]. |
| $T^{set}/\beta^{set}$ | Indoor temperature set-point/tolerance. |
| $P^{max}/P^{min}$ | Maximum/Minimum colling power. |
| $\boldsymbol{H}/\boldsymbol{h}$ | Coefficient matrix/vector for HVAC sets. |
| $\boldsymbol{H}^{base}/\boldsymbol{h}^{base}$ | Coefficient matrix/vector for base sets. |
| $\boldsymbol{\Gamma}^{aff}/\boldsymbol{\gamma}^{aff}$ | Affine matrix/Translation vector. |
| $C^f$ | Fuel cost for thermal units. |
| $C^{su}/C^{sd}$ | Start-up/shut-down price of thermal units. |
| $C^{hu}/C^{hd}$ | Up/down reserve capacity price for HVACs. |
| $C^{gud}/C^{gdd}$ | Up/down reserve deployment price for thermal units. |
| $C^{hud}/C^{hdd}$ | Up/down reserve deployment price for HVACs. |
| $C^{wc}$ | Penalty price for wind curtailment. |
| $C^{ls}$ | Penalty price for load shedding. |
| $P^{wd}/P^{ld}$ | Forecast wind power/load demand. |
| $P^{ref}$ | Base power for aggregated HVAC. |
| $S$ | Transformation factor on line. |
| $\overline{P}_l$ | Transmission line capacity. |
| $P(s)$ | Probability of each scenario |

*Decision Variables:*

| | |
|---|---|
| $T^a/T^m$ | Indoor air/mass temperature. |
| $P$ | Cooling power of single HVAC. |

This work was supported by the National Natural Science Foundation of China (52177088) and the Fundamental Research Funds for the Central Universities (YCJJ20230463). *(Corresponding author: Xiaomeng Ai).*

J. Liu, X. Ai, J. Fang, S. Cui, S. Wang, W. Yao and J. Wen are with the State Key Laboratory of Advanced Electromagnetic Technology, Huazhong University of Science and Technology, Wuhan 430074, China (e-mail: spencerplusmail@foxmail.com; xiaomengai@hust.edu.cn; jfa@hust.edu.cn; shichang_cui@hust.edu.cn; shengshiwang@hust.edu.cn; w.yao@hust.edu.cn; jinyu.wen@hust.edu.cn).

| | |
|---|---|
| $P^{app}$ | Approximate aggregated HVAC power. |
| $S^u/S^d$ | Indicator for the start-up/shut-down action of thermal units. |
| $P^{hu}/P^{hd}$ | Up/down reserve capacity for HVACs. |
| $P^{gud}/P^{gdd}$ | Up/down reserve deployment for thermal units. |
| $P^{hud}/P^{hdd}$ | Up/down reserve deployment for HVACs. |
| $P^{wc}$ | Power of wind curtailment. |
| $P^{ls}$ | Power of load shedding. |
| $P^g/S^g$ | Day-ahead generation/status of thermal units. |

# I. Introduction

With the increased renewables adding fast variations to power system, it requires more flexible resources for power system to follow the intra-hour changes in net load [1]. On the other hand, heating, ventilation, and air conditioning (HVAC) loads have been recognized as a promising kind of flexible resources for power system owing to their thermal inertia and rapid response capacities [2]. Thus, utilizing the intra-hour flexibility of HVACs for reserve provision can help the power system operator deal with the intermittency and uncertainty of renewables, thereby enhancing the economic and reliable operation in power system [3]. Since the capacity of single HVAC is quite limited, massive flexible HVACs are usually considered as a whole to provide the aggregated flexibility [4]. In this way, the HVAC aggregator serves as an interface between power system and HVACs who manages the contracted HVACs, estimates the aggregated flexibility of HVACs, and assigns power profiles to each HVAC [5]. With the participation of HVAC aggregators, the concept of hierarchical dispatch has been introduced for the efficient scheduling of HVACs to avoid heavy communication and computational burden [6].

However, the key challenge for HVAC aggregators is to estimate the aggregated HVAC flexibility in an accurate and reliable way [7]. The flexibility set of each HVAC can be described by a subset in the power space, and the exact aggregated flexibility refers the point-wise sum of these sets, i.e., the Minkowski sum (M-sum) of the HVAC feasible regions with heterogeneous parameters which is computationally intractable [8]. Thus, several attempts have been made to approximate the aggregated feasible region of HVACs, but two significant issues still remain.

The first issue is how to describe the intra-hour aggregated flexibility of massive HVACs, which is an extremely high-dimensional feasible region in continuous-time (CT) function space. Traditional works usually approximate the M-sum in discrete-time (DT) aggregation in low-dimensional feasible region, which can be divided into two types: outer-approximation and inner-approximation. The main idea of the former one is to approximate the aggregated feasible region via simple linear summations, such as the widely used virtual battery models which directly sum up the boundaries of each HVAC [3], [4], or use average parameters to represent the parameters of the aggregated flexibility [9]. The outer-approximation is computationally efficient but may contain undesired infeasible region, leading to disaggregation infeasibility. In contrast, the inner-approximation ensures no constraint violations in the disaggregation process. A widely used way is to utilize geometric techniques to transform a base set to approximate the exact feasible region, including box-based [10], zonotope-based [11], ellipsoid-based [12], and polytope-based [5], [8], [13] models. These geometric models can efficiently calculate the M-sum of the transformed base sets to approximate the exact aggregated flexibility of HVACs, among which the polytope-based model balances accuracy and computational complexity [5]. However, the existing polytope-based model has also been observed to reach an over-conservative aggregation result when there is great heterogeneity in high-dimensional feasible region due to its poor geometric adaptability [5].

Worse still, the feasible region of each HVAC is originally described by a set of partial differential equations in CT formulation [14], which is very high-dimensional in function space. Although the feasible region of each HVAC is usually discretized for simplification in power system dispatch, the intra-hour thermal dynamic is also neglected in existing DT aggregation works, making the aggregated flexibility of HVACs underestimated. On the other hand, with the increased renewables adding fast variations to power system, the intra-hour ramping events occur much more commonly [15], [16], making it an urgent need to enhance the intra-hour flexibility in power system. Thus, it requires further research to extend the traditional polytope-based aggregation model to a CT formulation, so as to provide the intra-hour aggregated flexibility of HVACs for power system while keeping high aggregation accuracy and relatively low computational complexity.

The second issue is how to efficiently deal with the impact of outdoor temperature uncertainty in HVAC aggregation. The outdoor temperature uncertainty has been observed to have a great impact on the feasible region of HVACs [17], while it is usually neglected in existing aggregation works. Ignoring the outdoor temperature uncertainty, the aggregators will miscalculate the aggregated flexibility of HVACs and cannot guarantee the disaggregation feasibility in practice [18]. To handle the outdoor temperature uncertainty, traditional methods include stochastic optimization [19], [20] and robust optimization [21], [22]. In the former one, a large number of scenarios are needed for better economy and will drastically increase the computation burden [23]. The latter one focuses on the worst scenarios with low occurrence probability and is over-conservative [18]. Hence, the outdoor temperature uncertainty is constructed by the distributionally robust chance constraints (DRCCs) in this paper, in view of their tractability and good out-of-sample performance [24]. However, how to efficiently incorporate the DRCCs into the CT aggregation model of HVACs still remains an open work.

To address the above research gaps, we first propose the CT aggregation model where the outdoor temperature uncertainty is considered for reliable aggregation. Then, we propose the reformulation techniques of the CT aggregation model for tractable calculation. Further, we propose a customized hierarchical dispatch framework via incorporating the CT aggregation model into reserve provision in power system dispatch. In the dispatch framework, the renewable uncertainty is considered for the economic dispatch of HVACs.

The main contributions are as follows, compared with other studies as shown in TABLE I.

TABLE I
COMPARISONS OF HVAC AGGREGATION MODELS.

| Reference | Intra-hour Flexibility | Aggregation Uncertainty | Aggregation Accuracy | Computational Complexity |
|---|---|---|---|---|
| [10] | No | No | Moderate | Low |
| [12] | No | No | High | High |
| [8] | No | No | Moderate | Low |
| [5], [7] | No | Yes | Moderate | Low |
| [3] | Yes | No | Low | Moderate |
| [4], [9], [11] | No | No | Low | Low |
| This paper | **Yes** | **Yes** | **High** | **Low** |

1) To the best of our knowledge, it is the first time to extend the traditional polytope-based aggregation model to a CT formulation, in order to reveal the intra-hour flexibility of HVACs. For accurate aggregation, a new affine transformation, with great geometric adaptability, is designed to handle the heterogeneity in high-dimensional feasible region. In addition, for further reliable aggregation in uncertain environment, the outdoor temperature uncertainty is constructed by DRCCs and integrated into the aggregation model.

2) For the tractable calculation of the proposed CT aggregation model, a cascade of tailored reformulation techniques is proposed. First, the Bernstein polynomial (BP) spline is used to transform the feasible region of the CT aggregation model into a finite-dimensional polytope in algebraic space. Next, the transformed polytope is projected into another polytope with the lower and full dimensional space to avoid missing any dimensions in aggregation. Further, the linearization transformation is explored. Finally, the CT aggregation model is transformed into linear programming that can be solved by off-the-shelf solvers.

3) Through extensive case studies, we verify that more potential reserve capacity of HVACs can be identified by our aggregation model in power system dispatch, thereby improving the operational security and economy against the renewable uncertainty. In addition, our aggregation model, with the controllable confidence level, provides a regulation measure for HVAC aggregators to balance the economy and risk attitude against the outdoor temperature uncertainty in real applications.

Following is the remainder of this paper. Section II describes the CT aggregation model of HVACs. Section III discusses the reformulation techniques of the CT aggregation model. Section IV presents the customized hierarchical dispatch framework, which incorporates the CT aggregation model into reserve provision in power system dispatch. Case studies are conducted in Section V. Section VI concludes this paper.

The logical relationships between the sections are shown in Fig. 1.

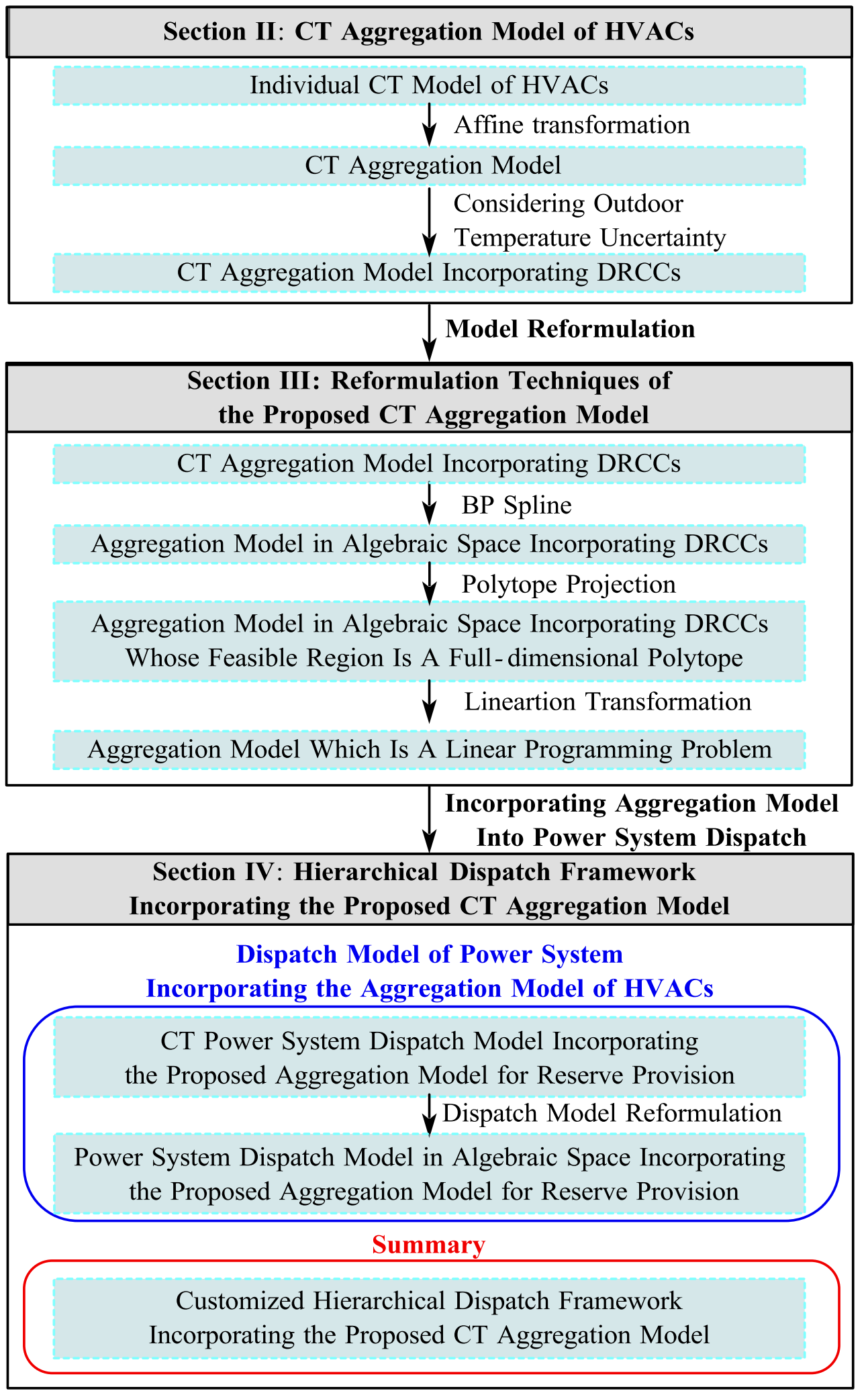

Fig. 1. Relationship of Section II-IV in this paper.

## II. CT AGGREGATION MODEL OF MASSIVE FLEXIBLE HVACS

This section first introduces the individual CT model of HVACs. Then, the CT aggregation model based on affine transformation is investigated. Next, the impact of outdoor temperature uncertainty on aggregation is considered.

### *A. Individual CT Model of HVACs*

We adopt the second-order equivalent thermal parameter model [14] as shown in (1.a)-(1.c), to depict the thermal dynamic of individual HVAC.

$$\frac{\mathrm{d}T_{i,k}^{a}(\tau)}{\mathrm{d}\tau}=a_{i,k}^{c1}T_{i,k}^{a}(\tau)+a_{i,k}^{c2}T_{i,k}^{m}(\tau)+a_{i,k}^{c3}T_{i,k}^{o}(\tau)+a_{i,k}^{c4}P_{i,k}(\tau) \quad (1.a)$$

$$\frac{\mathrm{d}T_{i,k}^{m}(\tau)}{\mathrm{d}\tau}=a_{i,k}^{c5}T_{i,k}^{a}(\tau)+a_{i,k}^{c6}T_{i,k}^{m}(\tau)+a_{i,k}^{c7}P_{i,k}(\tau) \quad (1.b)$$

$$\begin{cases} a_{i,k}^{c1} = -\left(H_{i,k}^{a} + H_{i,k}^{m}\right) / C_{i,k}^{a}, a_{i,k}^{c2} = -H_{i,k}^{m} / C_{i,k}^{a} \\ a_{i,k}^{c3} = -H_{i,k}^{a} / C_{i,k}^{a}, a_{i,k}^{c4} = -(1 - f_{i,k}^{h}) \mu_{i,k}^{h} / C_{i,k}^{a} \\ a_{i,k}^{c5} = H_{i,k}^{m} / C_{i,k}^{m}, a_{i,k}^{c6} = -H_{i,k}^{m} / C_{i,k}^{m} \\ a_{i,k}^{c7} = (1 - f_{i,k}^{h}) \mu_{i,k}^{h} / C_{i,k}^{m} \end{cases} \tag{1.c}$$

In addition, the HVAC system should include the indoor temperature limit constraint as shown in (1.d), which preserves the users' comfort.

$$T_{i,k}^{set} - \beta_{i,k}^{set} \le T_{i,k}^{a}(\tau) \le T_{i,k}^{set} + \beta_{i,k}^{set} \tag{1.d}$$

Also, the power limit constraint (1.e) is considered as below.

$$P_{i,k}^{\min} \le P_{i,k}(\tau) \le P_{i,k}^{\max} \tag{1.e}$$

Then, the feasible region of HVAC (1.a)-(1.e) can be expressed in the form of the H-representation of a CT polytope $\mathbb{U}_{i,k}$ as shown in (2). Note that the bold variable vectors, i.e., $\boldsymbol{P}_{i,k}$, $\boldsymbol{T}_{i,k}^{a}$, $\boldsymbol{T}_{i,k}^{m}$, $\frac{\mathrm{d}\boldsymbol{T}_{i,k}^{a}}{\mathrm{d}\tau}$, and $\frac{\mathrm{d}\boldsymbol{T}_{i,k}^{m}}{\mathrm{d}\tau}$, represent the CT sets of the corresponding italic variables.

$$\mathbb{U}_{i,k} = \{\boldsymbol{H}_{i,k}\left[\left(\boldsymbol{P}_{i,k}\right)^{\mathrm{T}}, \left(\boldsymbol{T}_{i,k}^{a}\right)^{\mathrm{T}}, \left(\boldsymbol{T}_{i,k}^{m}\right)^{\mathrm{T}}, \left(\frac{\mathrm{d}\boldsymbol{T}_{i,k}^{a}}{\mathrm{d}\tau}\right)^{\mathrm{T}}, \left(\frac{\mathrm{d}\boldsymbol{T}_{i,k}^{m}}{\mathrm{d}\tau}\right)^{\mathrm{T}}\right]^{\mathrm{T}} \le \boldsymbol{h}_{i,k}\} \tag{2}$$

*B. Aggregation Model Based on Affine Transformation*

For the aggregator at node $i$ who controls $N_i^K$ HVACs, its exact aggregated feasible region can be calculated via (3).

$$\mathbb{U}_{i}^{agg} = \biguplus_{k} \mathbb{U}_{i,k} = \left\{\boldsymbol{P}_{i}^{agg} = \sum_{k} \boldsymbol{P}_{i,k}, \boldsymbol{P}_{i,k} \in \mathbb{U}_{i,k}\right\} \tag{3}$$

where $\uplus$ is M-sum. $\boldsymbol{P}^{agg}$ is the variable vector of the aggregated power.

Since the exact M-sum of HVACs is computationally intractable, a novel inner-approximation model based on affine transformation is proposed in this paper. Specifically, a base polytope set is first constructed as shown in (4.a), where its coefficient matrices, i.e., $\boldsymbol{H}_i^{base}$ and $\boldsymbol{h}_i^{base}$, can be specified by averaging the parameters of all HVACs within the aggregator. Next, the matrices of affine transformation, i.e., the affine matrices $\boldsymbol{\Gamma}_{i,k}^{aff}$ and the translation vectors $\boldsymbol{\gamma}_{i,k}^{aff}$, are optimized to be selected so that the affine-transformed polytope $\mathbb{U}_{i,k}^{aff}$ can closely approximate the exact HVAC polytope $\mathbb{U}_{i,k}$ while the containment constraints of polytopes are respected as shown in (4.b). After all HVAC polytopes at node $i$ are approximated, the approximate M-sum of each $\mathbb{U}_{i,k}^{aff}$ can be calculated by (4.c). Since the exact M-sum of $\mathbb{U}_{i,k}^{aff}$ is computationally intractable, we provide an inner-approximation as shown in (4.c) for tractable calculation. The correctness of (4.c) is proven in Appendix A.

$$\mathbb{U}_{i}^{base} = \{\boldsymbol{H}_{i}^{base}\left[\left(\boldsymbol{P}_{i}\right)^{\mathrm{T}}, \left(\boldsymbol{T}_{i}^{a}\right)^{\mathrm{T}}, \left(\boldsymbol{T}_{i}^{m}\right)^{\mathrm{T}}, \left(\frac{\mathrm{d}\boldsymbol{T}_{i}^{a}}{\mathrm{d}\tau}\right)^{\mathrm{T}}, \left(\frac{\mathrm{d}\boldsymbol{T}_{i}^{m}}{\mathrm{d}\tau}\right)^{\mathrm{T}}\right]^{\mathrm{T}} \le \boldsymbol{h}_{i}^{base}\} \tag{4.a}$$

$$\mathbb{U}_{i,k}^{aff} = \boldsymbol{\Gamma}_{i,k}^{aff} \mathbb{U}_{i}^{base} + \boldsymbol{\gamma}_{i,k}^{aff} \subseteq \mathbb{U}_{i,k} \tag{4.b}$$

$$\mathbb{U}_{i}^{app} = \sum_{k}\left(\boldsymbol{\Gamma}_{i,k}^{aff}\right)\mathbb{U}_{i}^{base} + \sum_{k}\left(\boldsymbol{\gamma}_{i,k}^{aff}\right) \subseteq \biguplus_{k} \mathbb{U}_{i,k}^{aff} \subseteq \mathbb{U}_{i}^{agg} \tag{4.c}$$

To illustrate the advantage of the proposed affine transformation, we show a 2-D example in Fig. 2. The base set is set as a square while the exact polytope is a diamond. Following the conventional aggregation model which sets $\boldsymbol{\Gamma}_{i,k}^{aff}$ as a constant (the scaling in all dimensions is uniform), the final approximate polytope is still a square while losing much feasible region. Indeed, this model is restricted by its essence of structural preservation with poor geometric adaptability, and may be over-conservative in high-dimensional feasible region. If we assume the scaling in all dimensions can be variable ($\boldsymbol{\Gamma}_{i,k}^{aff}$ is set as a diagonal matrix), the final approximate polytope becomes a rectangle with higher accuracy than the square. However, this model is also restricted by the limited transformation techniques (non-uniform scaling and translation only). Further, if we extend the transformation techniques into the general affine transformation ($\boldsymbol{\Gamma}_{i,k}^{aff}$ is set as a general matrix), the final approximate polytope becomes an exact diamond with the highest accuracy thanks to its great geometric adaptability. The above results show that our model enjoys great geometric adaptability, being applicable to approximate the high-dimensional feasible region.

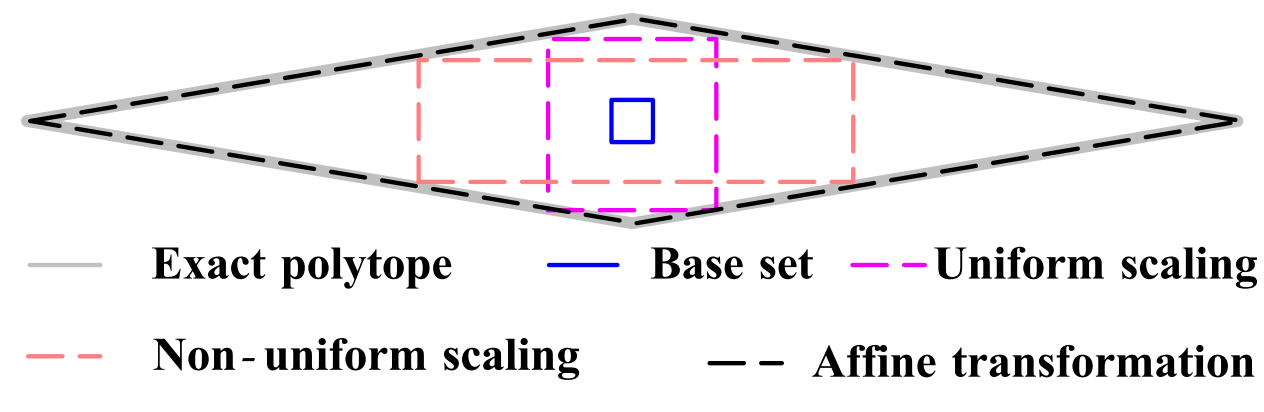

Fig. 2. Diagram of the inner-approximation.

*C. Impact of with Outdoor Temperature Uncertainty*

In (1), the feasible region of each HVAC is determined by a series of parameters, among which the outdoor temperature $T_{i,k}^{o}(\tau)$ is uncertain parameters obtained by prediction. It varies in real-time dispatch and has a great impact on the feasible region of HVAC, thereby being integrated into the modeling. Since $T_{i,k}^{o}(\tau)$ is manifested as the uncertainty of the vector $\boldsymbol{h}_{i,k}$, (4.b) can be recast into the following DRCCs (5):

$$\inf_{f(\boldsymbol{h}_{i,k}) \in \mathcal{D}} \mathbb{P}_{\boldsymbol{h}_{i,k}} \left\{\boldsymbol{\Gamma}_{i,k}^{aff} \mathbb{U}_{i}^{base} + \boldsymbol{\gamma}_{i,k}^{aff} \subseteq \mathbb{U}_{i,k}\right\} \ge 1 - \varepsilon \tag{5}$$

where $\varepsilon$ is the allowable violation probability. $f(\boldsymbol{h}_{i,k})$ is the probability distribution function of $\boldsymbol{h}_{i,k}$. The ambiguity set $\mathcal{D}$ is defined as in (6) where $\mathbb{E}$ is the expectation operator, $\boldsymbol{\mu}_{i,k}$ is empirical mean vector, and $\boldsymbol{\sigma}_{i,k}$ is empirical covariance matrix.

$$\mathcal{D} = \left\{\mathbb{E}\left[\boldsymbol{h}_{i,k}\right] = \boldsymbol{\mu}_{i,k}, \mathbb{E}_{i,k}\left[\left(\boldsymbol{h}_{i,k} - \boldsymbol{\mu}_{i,k}\right)\left(\boldsymbol{h}_{i,k} - \boldsymbol{\mu}_{i,k}\right)^{\mathrm{T}}\right] = \boldsymbol{\sigma}_{i,k}\boldsymbol{\sigma}_{i,k}^{\mathrm{T}}\right\} \tag{6}$$

To guarantee the maximum inner-approximation region under the outdoor temperature uncertainty, the following problem is proposed for each HVAC to determine the optimal value of $\boldsymbol{\Gamma}_{i,k}^{aff}$ and $\boldsymbol{\gamma}_{i,k}^{aff}$:

$$\max_{\boldsymbol{\Gamma}_{i,k}^{aff},\boldsymbol{\gamma}_{i,k}^{aff}} |\det(\boldsymbol{\Gamma}_{i,k}^{aff})| \tag{7.a}$$

$$\text{s.t.} \inf_{f(\boldsymbol{h}_{i,k})\in\mathcal{D}} \mathbb{P}_{\boldsymbol{h}_{i,k}} \left\{ \boldsymbol{\Gamma}_{i,k}^{aff} \mathbb{U}_i^{base} + \boldsymbol{\gamma}_{i,k}^{aff} \subseteq \mathbb{U}_{i,k} \right\} \geq 1-\varepsilon \tag{7.b}$$

where the objective function is to maximize the absolute determinant of $\boldsymbol{\Gamma}_{i,k}^{aff}$. Since the volume of high-dimensional $\mathbb{U}_{i,k}^{aff}$ can be obtained via multiplying $|\det(\Gamma_{i,k}^{aff})|$ by the volume of $\mathbb{U}_i^{base}$, maximizing the volume of $\mathbb{U}_{i,k}^{aff}$ is equivalent to maximizing $|\det(\Gamma_{i,k}^{aff})|$.

To sum up, the overall aggregation process is presented in Fig. 3. The aggregator at node $i$ first construct the base set (4.a). Then, the feasible regions of each HVAC within the aggregator are approximated in a parallel manner via solving (7) to determine the optimal value of $\boldsymbol{\Gamma}_{i,k}^{aff}$ and $\boldsymbol{\gamma}_{i,k}^{aff}$ in (4.b). Next, the inner-approximation feasible region of HVACs within the aggregator can be obtain via (4.c)

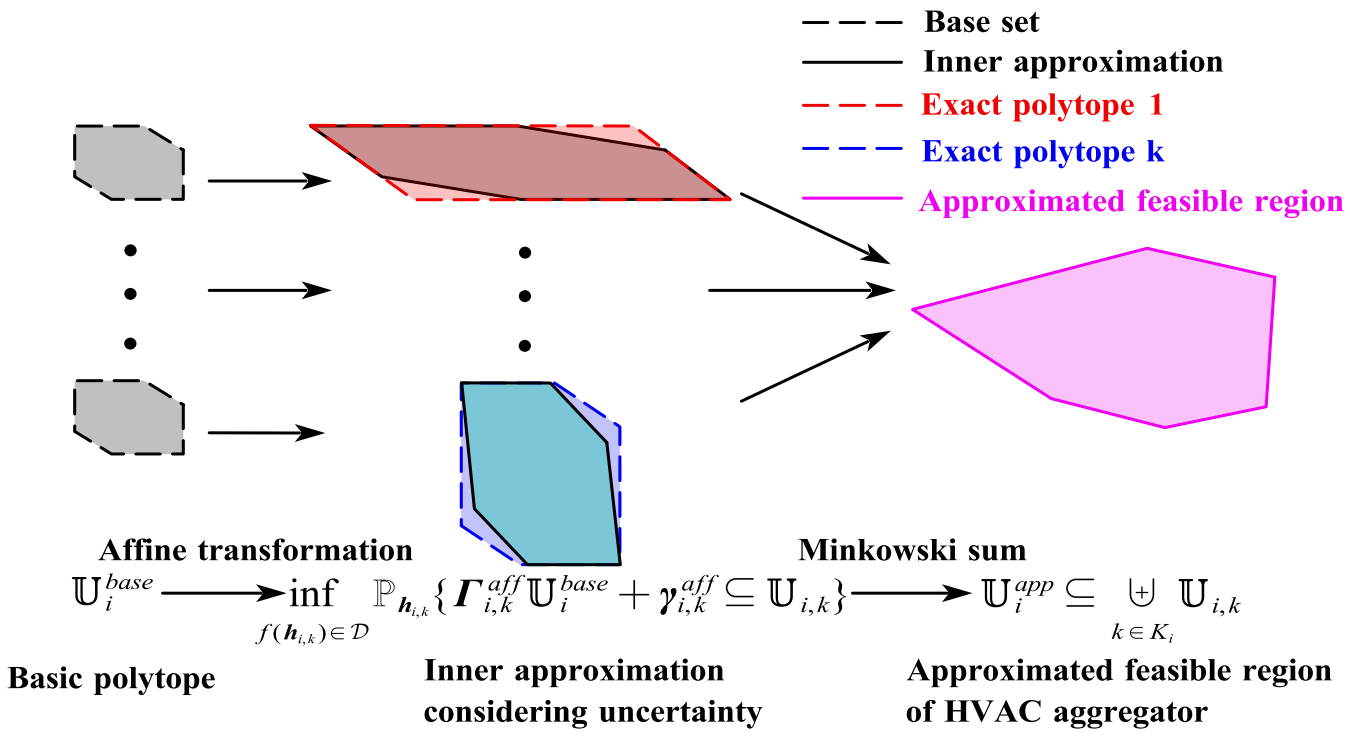

Fig. 3. Diagram of the aggregation process.

However, the problem (7) is CT optimization in function space with nonlinear constraints and objective function. Thus, reformulation techniques are required to efficiently solve (7).

## III. Reformulation Techniques for the Proposed CT Aggregation Model

To tractably calculate the proposed CT aggregation model, this section develops a cascade of tailored reformulation techniques, including BP spline, polytope projection, and linearization transformation.

### A. BP Spline

BP spline [25] is first used to transform the original CT model of HVACs in function space (1) into a finite-dimensional polytope in algebraic space. Specifically, the dispatch horizon is subdivided into $N^T$ periods with the same length of $\Delta t$ (typically 1 hour for day-ahead dispatch). For each period, the CT function $F(\tau)$ is mapped into the cubic BPs as shown in (8) where $B_{3,q}(\tau) = \binom{3}{q}\tau^q(1-\tau)^{3-q}$ for $\tau \in [0,1]$ and $q = 0,1,2,3$ is a cubic BP, and $\boldsymbol{F^B} = [F_0^B, F_1^B, F_2^B, F_3^B]^\mathrm{T}$ are spline coefficients.

$$F(\tau) = \sum_{q=0}^{3} F_q^B B_{3,q}(\tau) = \left(\boldsymbol{F^B}\right)^\mathrm{T} \boldsymbol{B}_3(\tau), \tau \in [0,1] \tag{8}$$

The BPs are utilized based on their favorable properties as shown in (9) where $\boldsymbol{B}_2(\tau)$ represents quadratic BPs, and $\boldsymbol{W}, \boldsymbol{J}$, and $\boldsymbol{L}$ are the operation matrices of known entries [25], [26].

$$\begin{cases} \dfrac{\mathrm{d}F(\tau)}{\mathrm{d}\tau} = \dfrac{\mathrm{d}\left(\boldsymbol{F^B}\right)^\mathrm{T} \boldsymbol{B}_3(\tau)}{\mathrm{d}\tau} = \left(\boldsymbol{W}\boldsymbol{F^B}\right)^\mathrm{T} \boldsymbol{B}_2(\tau) \\ \int_0^1 F(\tau)\mathrm{d}\tau = 1^\mathrm{T}\boldsymbol{F^B}/4 \\ \int_0^\tau F(\tau)\mathrm{d}\tau = \int_0^\tau \left(\boldsymbol{F^B}\right)^\mathrm{T} \boldsymbol{B}_3(\tau)\mathrm{d}\tau \cong \left(\boldsymbol{F^B}\right)^\mathrm{T} \boldsymbol{L}\boldsymbol{B}_3(\tau) \\ F(\tau) = 0 \Leftrightarrow \left(\boldsymbol{F^B}\right)^\mathrm{T} \boldsymbol{B}_3(\tau) = 0 \Leftrightarrow \boldsymbol{F^B} = 0 \\ F(\tau) \leq 0 \Leftrightarrow \left(\boldsymbol{F^B}\right)^\mathrm{T} \boldsymbol{B}_3(\tau) \leq 0 \Leftarrow \boldsymbol{J}\boldsymbol{F^B} \leq 0 \end{cases} \tag{9}$$

With the above properties, we can reformulate the integral terms, differential terms, equality terms and inequality terms in CT feasible region (1) as (10.a)-(10.e):

$$\begin{cases} \boldsymbol{T}_{i,k,t}^{a,B} - \boldsymbol{T}_{i,k,t}^{a,B,ini} = \boldsymbol{L}^T(a_{i,k}^{c1}\boldsymbol{T}_{i,k,t}^{a,B} + a_{i,k}^{c2}\boldsymbol{T}_{i,k,t}^{m,B} + a_{i,k}^{c3}\boldsymbol{T}_{i,k,t}^{o,B} + a_{i,k}^{c4}\boldsymbol{P}_{i,k,t}^{B}) \\ T_{i,k,t,q}^{a,B,ini}\big|_{t=1} = T_{i,k}^{set}, T_{i,k,t,q}^{a,B,ini}\big|_{t>1} = T_{i,k,t-1,3}^{a,B}, q = 0,1,2,3 \end{cases} \tag{10.a}$$

$$\begin{cases} \boldsymbol{T}_{i,k,t}^{m,B} - \boldsymbol{T}_{i,k,t}^{m,B,ini} = \boldsymbol{L}^T(a_{i,k}^{c5}\boldsymbol{T}_{i,k,t}^{a,B} + a_{i,k}^{c6}\boldsymbol{T}_{i,k,t}^{m,B} + a_{i,k}^{c7}\boldsymbol{P}_{i,k,t}^{B}) \\ T_{i,k,t,q}^{m,B,ini}\big|_{t=1} = T_{i,k}^{m0}, T_{i,k,t,q}^{m,B,ini}\big|_{t>1} = T_{i,k,t-1,3}^{m,B,ini}, q = 0,1,2,3 \end{cases} \tag{10.b}$$

$$T_{i,k}^{set} - \beta_{i,k}^{set} \leq \boldsymbol{J}\boldsymbol{T}_{i,k,t}^{a,B} \leq T_{i,k}^{set} + \beta_{i,k}^{set} \tag{10.c}$$

$$P_{i,k}^{\min} \leq \boldsymbol{J}\boldsymbol{P}_{i,k,t}^{B} \leq P_{i,k}^{\max} \tag{10.d}$$

$$P_{i,k,t,3}^{B} = P_{i,k,t+1,0}^{B}, P_{i,k,t,3}^{B} - P_{i,k,t,2}^{B} = P_{i,k,t+1,1}^{B} - P_{i,k,t+1,0}^{B} \tag{10.e}$$

where $[\cdot]_t^B \in \mathbb{R}^4$ is the spline coefficient vector of $[\cdot](\tau)$ in period $t$. (10.a)-(10.d) are the corresponding representation of (1.a)-(1.e) after BP splines. (10.e) is the first-order continuity constraint for the smooth connection between adjacent periods.

### B. Polytope Projection

After the BP spline, the original CT model of HVACs is recast as a finite-dimensional polytope in algebraic space. However, due to the existence of equality constraints (10.a), (10.b) and (10.e), the derived polytope is not a full-dimensional polytope [5]. Thus, there will be some dimensional missing in aggregation, making the objective function $\det(\Gamma_{i,k}^{aff})$ unbounded. To avoid this, the polytope projection is proposed herein to obtain an equivalent full-dimensional polytope in power space. Our main goal is to eliminate redundant variables based on the equality constraints.

First, we eliminate $\boldsymbol{T}_{i,k}^{m,B}$ and $\boldsymbol{T}_{i,k}^{a,B}$. (10.a) and (10.b) can be rewritten as the compact form as following:

$$\begin{cases} \boldsymbol{A}_{i,k}^{c1}\boldsymbol{T}_{i,k}^{m,B} + \boldsymbol{A}_{i,k}^{c2}\boldsymbol{T}_{i,k}^{a,B} + \boldsymbol{A}_{i,k}^{c3}\boldsymbol{P}_{i,k}^{B} = \boldsymbol{A}_{i,k}^{c4} \\ \boldsymbol{A}_{i,k}^{c5}\boldsymbol{T}_{i,k}^{m,B} + \boldsymbol{A}_{i,k}^{c6}\boldsymbol{T}_{i,k}^{a,B} + \boldsymbol{A}_{i,k}^{c7}\boldsymbol{P}_{i,k}^{B} = \boldsymbol{A}_{i,k}^{c8} \end{cases} \tag{11.a}$$

where $\boldsymbol{A}_{i,k}^{[\cdot]}$ is the compact coefficient matrix, and $[\cdot]_{i,k}^B \in \mathbb{R}^{4N^T}$ is the spline coefficient vector of $[\cdot]_{i,k,t}^B$. Since $\boldsymbol{A}_{i,k}^{c1}$, $\boldsymbol{A}_{i,k}^{c2}$, $\boldsymbol{A}_{i,k}^{c5}$, and $\boldsymbol{A}_{i,k}^{c6}$ are invertible matrix, we recast (10.c) as (11.b).

$$T_{i,k}^{set} - \beta_{i,k}^{set} \leq \boldsymbol{J}(\boldsymbol{A}_{i,k}^{c10} - \boldsymbol{A}_{i,k}^{c9}\boldsymbol{P}_{i,k}^{B}) \leq T_{i,k}^{set} + \beta_{i,k}^{set} \tag{11.b}$$

where

$$\begin{cases} \boldsymbol{A}_{i,k}^{c9} = \boldsymbol{A}_{i,k}^{c11}\left(\boldsymbol{A}_{i,k}^{c7} - \boldsymbol{A}_{i,k}^{c5}\left(\boldsymbol{A}_{i,k}^{c1}\right)^{-1}\boldsymbol{A}_{i,k}^{c3}\right) \\ \boldsymbol{A}_{i,k}^{c10} = \boldsymbol{A}_{i,k}^{c11}\left(\boldsymbol{A}_{i,k}^{c8} - \boldsymbol{A}_{i,k}^{c5}\left(\boldsymbol{A}_{i,k}^{c1}\right)^{-1}\boldsymbol{A}_{i,k}^{c4}\right) \\ \boldsymbol{A}_{i,k}^{c11} = \left(\boldsymbol{A}_{i,k}^{c6} - \boldsymbol{A}_{i,k}^{c5}\left(\boldsymbol{A}_{i,k}^{c1}\right)^{-1}\boldsymbol{A}_{i,k}^{c2}\right)^{-1} \end{cases} \quad (11.c)$$

Then $\boldsymbol{T}_{i,k}^{m,B}$ and $\boldsymbol{T}_{i,k}^{a,B}$ are eliminated.

Next, we reconstruct $\boldsymbol{P}_{i,k}^{a,B}$ by recasting (10.e) as below:

$$\boldsymbol{P}_{i,k}^{B} = \boldsymbol{D}\boldsymbol{P}_{i,k}^{BB} \quad (11.d)$$

where $\boldsymbol{P}_{i,k}^{BB} \in \mathbb{R}^{2N^T+4}$ is the projected variable vector, and $\boldsymbol{D} \in \mathbb{R}^{4N^T*(2N^T+4)}$ is the projection operation matrix. By substituting (11.d) into (11.b) and (10.d), we finally obtain the projected full-dimensional polytope of HVAC model as below:

$$\begin{cases} T_{i,k}^{set} - \beta_{i,k}^{set} \le \boldsymbol{J}(\boldsymbol{A}_{i,k}^{c10} - \boldsymbol{A}_{i,k}^{c9}\boldsymbol{D}\boldsymbol{P}_{i,k}^{BB}) \le T_{i,k}^{set} + \beta_{i,k}^{set} \\ P_{i,k}^{\min} \le \boldsymbol{D}\boldsymbol{P}_{i,k}^{BB} \le P_{i,k}^{\max} \end{cases} \quad (12)$$

Further, we can recast the original HVAC polytope (2) as the full-dimensional polytope as shown in (13). Also, we recast the original HVAC base set (4.a) as the full-dimensional polytope as shown in (14). Herein, $\boldsymbol{H}_{i,k}^{BB}$, $\boldsymbol{h}_{i,k}^{BB}$, $\boldsymbol{H}_{i}^{base,BB}$, and $\boldsymbol{h}_{i}^{base,BB}$ are corresponding compact coefficient matrices.

$$\mathbb{U}_{i,k} = \{\boldsymbol{H}_{i,k}^{BB}\boldsymbol{P}_{i,k}^{BB} \le \boldsymbol{h}_{i,k}^{BB}\} \quad (13)$$

$$\mathbb{U}_{i}^{base} = \{\boldsymbol{H}_{i}^{base,BB}\boldsymbol{P}_{i,k}^{BB} \le \boldsymbol{h}_{i}^{base,BB}\} \quad (14)$$

After the above projection, the original problem (7) can be recast as (15) where the issue of the dimensional missing is avoided.

$$\max_{\boldsymbol{\Gamma}_{i,k}^{aff},\boldsymbol{\gamma}_{i,k}^{aff}} |\det\left(\boldsymbol{\Gamma}_{i,k}^{aff,BB}\right)| \quad (15.a)$$

$$\text{s.t.} \inf_{f\left(\boldsymbol{h}_{i,k}^{BB}\right)\in\mathcal{D}} \mathbb{P}_{\boldsymbol{h}_{i,k}^{BB}}\left\{\boldsymbol{\Gamma}_{i,k}^{aff,BB}\boldsymbol{P}_{i,k}^{BB} + \boldsymbol{\gamma}_{i,k}^{aff,BB} \subseteq \mathbb{U}_{i,k}\right\} \ge 1-\varepsilon, \forall \boldsymbol{P}_{i,k}^{BB} \in \mathbb{U}_{i}^{base} \quad (15.b)$$

### C. Linearization Transformation

After the projection, there are some nonlinear terms in (15), including $\det(\Gamma_{i,k}^{aff,BB})$ in (15.a), the containment constraint in (15.b), and DRCCs in (15.b), which may cause computational intractability. Thus, linearization transformation is then proposed.

*1) Objective Function Transformation:* We approximate $|\det(\Gamma_{i,k}^{aff,BB})|$ with its first-order Taylor expansion as below, i.e., the trace of $\Gamma_{i,k}^{aff,BB}$, considering its satisfactory accuracy and computation tractability [27].

$$\max_{\boldsymbol{\Gamma}_{i,k}^{aff},\boldsymbol{\gamma}_{i,k}^{aff}} \operatorname{Tr}\left(\boldsymbol{\Gamma}_{i,k}^{aff,BB}\right) \quad (16)$$

*2) Containment Constraint Transformation:* The containment constraint $\boldsymbol{\Gamma}_{i,k}^{aff,BB}\boldsymbol{P}_{i,k}^{BB} + \boldsymbol{\gamma}_{i,k}^{aff,BB} \subseteq \mathbb{U}_{i,k}$ can be equivalently recast as in (17). The key method is to apply the Farkas' Lemma [28]. We prove the correctness of containment constraint reformulation in Appendix B.

$$\boldsymbol{\Lambda}_{i,k} \ge 0 \quad (17.a)$$

$$\boldsymbol{\Lambda}_{i,k}\boldsymbol{H}_{i}^{base,BB} = \boldsymbol{H}_{i,k}^{BB}\boldsymbol{\Gamma}_{i,k}^{aff,BB} \quad (17.b)$$

$$\boldsymbol{\Lambda}_{i,k}\boldsymbol{h}_{i}^{base,BB} + \boldsymbol{H}_{i,k}^{BB}\boldsymbol{\gamma}_{i,k}^{aff,BB} \le \boldsymbol{h}_{i,k}^{BB} \quad (17.c)$$

where $\boldsymbol{\Lambda}_{i,k}$ is the auxiliary variable matrix.

*3) Transformation of DRCCs:* After the reformulation of containment constraints, the uncertain vector $\boldsymbol{h}_{i,k}^{BB}$ only occurs in (17.c). Thus, the original DRCCs (15.b) can be equivalent to (18). Then, we reformulate (18) as (19). We prove the correctness of the reformulation of DRCCs in Appendix C.

$$\inf_{f(\boldsymbol{h}_{i,k}^{BB})\in\mathcal{D}} \mathbb{P}_{\boldsymbol{h}_{i,k}^{BB}}\left\{\boldsymbol{\Lambda}_{i,k}\boldsymbol{h}_{i}^{base,BB} + \boldsymbol{H}_{i,k}^{BB}\boldsymbol{\gamma}_{i,k}^{aff,BB} \le \boldsymbol{h}_{i,k}^{BB}\right\} \ge 1-\varepsilon \quad (18)$$

$$\begin{cases} \boldsymbol{\Lambda}_{i,k}(m,:)\boldsymbol{h}_{i}^{base,BB} + \boldsymbol{H}_{i,k}^{BB}(m,:)\boldsymbol{\gamma}_{i,k}^{aff,BB} \le \boldsymbol{h}_{i,k}^{BB}(m), m \in M_d \\ \boldsymbol{\Lambda}_{i,k}(m,:)\boldsymbol{h}_{i}^{base,BB} + \boldsymbol{H}_{i,k}^{BB}(m,:)\boldsymbol{\gamma}_{i,k}^{aff,BB} \\ \le \boldsymbol{\mu}_{i,k}(m) - \boldsymbol{\sigma}_{i,k}(m)\sqrt{\dfrac{1-\varepsilon_m}{\varepsilon_m}}, m \in M_u \end{cases} \quad (19)$$

where $m$ is row index, $M_u / M_d$ is the set of rows in $\boldsymbol{h}_{i,k}^{BB}$ without/with uncertain parameters respectively. $\varepsilon_m$ is the violation probability of the constraint at row $m$.

With the above linearization, the final aggregation model is given in (20) which is a linear programming problem, thereby being tractably solved by commercial solvers.

$$\max_{\boldsymbol{\Gamma}_{i,k}^{aff},\boldsymbol{\gamma}_{i,k}^{aff},\boldsymbol{\Lambda}_{i,k}} \operatorname{Tr}\left(\boldsymbol{\Gamma}_{i,k}^{aff,BB}\right) \quad (20.a)$$

$$\text{s.t. } (17.a),(17.b),(19) \quad (20.b)$$

## IV. Hierarchical Dispatch Framework Incorporating the Proposed CT Aggregation Model

This section introduces the customized hierarchical dispatch framework in power system which incorporates the CT aggregation model of HVACs into reserve provision to cope with the renewable uncertainty.

### A. Power System Dispatch Model Incorporating Aggregated HVAC Reserve

Based on the proposed CT aggregation model, the power system dispatch incorporating aggregated HVAC reserve is formulated as a two-stage CT stochastic optimization problem as shown in (21). The first stage is the day-ahead schedule of aggregated HVAC reserve capacity and unit commitment scheme. The second stage is the scenario-based intra-day regulation where the reserve from thermal units and aggregated HVAC are deployed.

$$\begin{aligned} \min \int_0^{N^T} \sum_i \begin{bmatrix} C_i^f(\tau) + C_i^{su}S_i^u(\tau) + C_i^{sd}S_i^d(\tau) \\ +C_i^{hu}P_i^{hu}(\tau) + C_i^{hd}P_i^{hd}(\tau) \end{bmatrix} \mathrm{d}\tau \\ + \int_0^{N^T} \sum_s \sum_i \begin{bmatrix} C_i^{gud}P_{i,s}^{gud}(\tau) + C_i^{gdd}P_{i,s}^{gdd}(\tau) \\ +C_i^{hud}P_{i,s}^{hud}(\tau) + C_i^{hgd}P_{i,s}^{hdd}(\tau) \\ +C_i^{wc}P_{i,s}^{wc}(\tau) + C_i^{ls}P_{i,s}^{ls}(\tau) \end{bmatrix} P(s)\mathrm{d}\tau \end{aligned} \quad (21.a)$$

$$\text{s.t.} \sum_i \left(P_i^g(\tau) + P_i^{wd}(\tau)\right) = \sum_i \left(P_i^{ld}(\tau) + P_i^{ref}(\tau)\right) \quad (21.b)$$

$$\left|\sum_i S_{ji}\left(P_i^g(\tau) + P_i^{wd}(\tau) - P_i^{ld}(\tau) - P_i^{ref}(\tau)\right)\right| \le \overline{P_{l,j}} \quad (21.c)$$

$$G\left(S_i^u(\tau), S_i^d(\tau), S_i^g(\tau), P_i^g(\tau)\right) \le 0 \quad (21.d)$$

$$\sum_i \left( P_i^g(\tau) + P_{i,s}^{gud}(\tau) - P_{i,s}^{gdd}(\tau) + P_{i,s}^{wd}(\tau) - P_{i,s}^{wc}(\tau) \right) = \sum_i \left( P_{i,s}^{ld}(\tau) - P_{i,s}^{ls}(\tau) + P_i^{ref}(\tau) + P_{i,s}^{hud}(\tau) - P_{i,s}^{hdd}(\tau) \right) \tag{21.e}$$

$$\left| \sum_i S_{ji} \begin{pmatrix} P_i^g(\tau) + P_{i,s}^{gud}(\tau) - P_{i,s}^{gdd}(\tau) + P_{i,s}^{wd}(\tau) - P_{i,s}^{wc}(\tau) \\ -(P_{i,s}^{ld}(\tau) - P_{i,s}^{ls}(\tau) + P_i^{ref}(\tau) + P_{i,s}^{hud}(\tau) - P_{i,s}^{hdd}(\tau)) \end{pmatrix} \right| \le \overline{P_{l,j}} \tag{21.f}$$

$$0 \le P_{i,s}^{wc}(\tau) \le P_{i,s}^{wd}(\tau), 0 \le P_{i,s}^{ls}(\tau) \le P_{i,s}^{ld}(\tau) \tag{21.g}$$

$$G\left( P_{i,s}^g(\tau), P_{i,s}^{gud}(\tau), P_{i,s}^{gud}(\tau) \right) \le 0 \tag{21.h}$$

$$\begin{cases} P_{i,s}^{app}(\tau) = P_i^{ref}(\tau) + P_{i,s}^{hud}(\tau) - P_{i,s}^{hdd}(\tau), P_{i,s}^{app}(\tau) \in \mathbb{U}_i^{app} \\ 0 \le P_{i,s}^{hud}(\tau) \le P_i^{hu}(\tau), 0 \le P_{i,s}^{hud}(\tau) \le P_i^{hd}(\tau) \end{cases} \tag{21.i}$$

The objective (21.a) is to minimize the total operational cost, including the first stage fuel cost of thermal units, startup/shutdown cost of thermal units, and HVAC reserve capacity costs, and the expected second stage upward/downward regulation cost of thermal units and HVACs, wind curtailment penalty, and load shedding penalty. The piecewise linear function of fuel cost for thermal units can be obtained from [25].

As for the first stage constraints, equation (21.b) is the first stage power balance constraint. Equation (21.c) is the first stage transmission line capacity limits. Equation (21.d) is the first stage constraints for thermal units, including minimum online and offline time constraints, generation range constraints, and ramping constraints [15].

As for the second stage constraints, equation (21.e) is the second stage power re-balance. Equation (21.f) is the second stage transmission line capacity limits. Equation (21.g) is the wind curtailment and load shedding limits. Equation (21.h) includes the second stage constraints for thermal units, i.e., up/down reserve deployment constraints, ramping constraints, and generation constraints [15]. Equation (21.i) includes the second stage constraints for HVACs, i.e., up/down reserve deployment and approximate polytope constraints.

### *B. Dispatch Model Reformulation*

Following the reformulation techniques introduced in Section III-A, we adopt the BP spline to shift the CT optimization problem (21) from function space to algebraic space as below:

$$\min \sum_i \sum_t \begin{pmatrix} C_i^{su} S_{i,t}^u + C_i^{sd} S_{i,t}^d + \mathbf{1}^{\mathrm{T}} \boldsymbol{C}_{i,t}^{f,B} / 4 \\ +(C_i^{hu} \mathbf{1}^{\mathrm{T}} \boldsymbol{P}_{i,t}^{hu,B} + C_i^{hd} \mathbf{1}^{\mathrm{T}} \boldsymbol{P}_{i,t}^{hd,B}) / 4 \end{pmatrix} + \sum_s \sum_i \sum_t \begin{pmatrix} (C_i^{gud} \mathbf{1}^{\mathrm{T}} \boldsymbol{P}_{i,t,s}^{gud,B} + C_i^{gdd} \mathbf{1}^{\mathrm{T}} \boldsymbol{P}_{i,t,s}^{gdd,B} \\ +C_i^{hud} \mathbf{1}^{\mathrm{T}} \boldsymbol{P}_{i,t,s}^{hud,B} + C_i^{hdd} \mathbf{1}^{\mathrm{T}} \boldsymbol{P}_{i,t,s}^{hdd,B} \\ +C_i^{wc} \mathbf{1}^{\mathrm{T}} \boldsymbol{P}_{i,t,s}^{wc,B} + C_i^{ls} \mathbf{1}^{\mathrm{T}} \boldsymbol{P}_{i,t,s}^{ls,B}) P(s) / 4 \end{pmatrix} \tag{22.a}$$

$$\text{s.t.} \ \sum_i \left( \boldsymbol{P}_{i,t}^{g,B} + \boldsymbol{P}_{i,t}^{wd,B} \right) = \sum_i \left( \boldsymbol{P}_{i,t}^{ld,B} + \boldsymbol{P}_{i,t}^{ref,B} \right) \tag{22.b}$$

$$\left| \sum_i S_{ji} \boldsymbol{J} \left( \boldsymbol{P}_{i,t}^{g,B} + \boldsymbol{P}_{i,t}^{wd,B} - \boldsymbol{P}_{i,t}^{ld,B} - \boldsymbol{P}_{i,t}^{ref,B} \right) \right| \le \overline{P_{l,j}} \tag{22.c}$$

$$G\left( S_{i,t}^u, S_{i,t}^d, S_{i,t}^g, \boldsymbol{P}_{i,t}^{g,B} \right) \le 0 \tag{22.d}$$

$$\sum_i \left( \boldsymbol{P}_{i,t}^{g,B} + \boldsymbol{P}_{i,t,s}^{gud,B} - \boldsymbol{P}_{i,t,s}^{gdd,B} + \boldsymbol{P}_{i,t,s}^{wd,B} - \boldsymbol{P}_{i,t,s}^{wc,B} \right) = \sum_i \left( \boldsymbol{P}_{i,t,s}^{ld,B} - \boldsymbol{P}_{i,t,s}^{ls,B} + \boldsymbol{P}_{i,t}^{ref,B} + \boldsymbol{P}_{i,t,s}^{hud,B} - \boldsymbol{P}_{i,t,s}^{hdd,B} \right) \tag{22.e}$$

$$\left| \sum_i S_{ji} \boldsymbol{J} \begin{pmatrix} \boldsymbol{P}_{i,t}^{g,B} + \boldsymbol{P}_{i,t,s}^{gud,B} - \boldsymbol{P}_{i,t,s}^{gdd,B} + \boldsymbol{P}_{i,t,s}^{wd,B} - \boldsymbol{P}_{i,t,s}^{wc,B} \\ -(\boldsymbol{P}_{i,t,s}^{ld,B} - \boldsymbol{P}_{i,t,s}^{ls,B} + \boldsymbol{P}_{i,t}^{ref,B} + \boldsymbol{P}_{i,t,s}^{hud,B} - \boldsymbol{P}_{i,t,s}^{hdd,B}) \end{pmatrix} \right| \le \overline{P_{l,j}} \tag{22.f}$$

$$0 \le \boldsymbol{J}\boldsymbol{P}_{i,s,t}^{wc,B} \le \boldsymbol{J}\boldsymbol{P}_{i,s,t}^{wd,B}, 0 \le \boldsymbol{J}\boldsymbol{P}_{i,s,t}^{ls,B} \le \boldsymbol{P}_{i,s,t}^{ld,B} \tag{22.g}$$

$$G\left( \boldsymbol{P}_{i,t,s}^{g,B}, \boldsymbol{P}_{i,t,s}^{gud,B}, \boldsymbol{P}_{i,t,s}^{gud,B} \right) \le 0 \tag{22.h}$$

$$\begin{cases} \boldsymbol{D}\boldsymbol{P}_{i,s,t}^{app,BB} = \boldsymbol{P}_{i,t}^{ref,B} + \boldsymbol{P}_{i,s,t}^{hud,B} - \boldsymbol{P}_{i,s,t}^{hdd,B}, \boldsymbol{P}_{i,s,t}^{app,BB} \in \mathbb{U}_i^{app} \\ 0 \le \boldsymbol{J}\boldsymbol{P}_{i,s,t}^{hud,B} \le \boldsymbol{P}_{i,t}^{hu,B}, 0 \le \boldsymbol{J}\boldsymbol{P}_{i,s,t}^{hud,B} \le \boldsymbol{P}_{i,t}^{hd,B} \end{cases} \tag{22.i}$$

$$[\cdot]_{i,k,t,3}^B = [\cdot]_{i,k,t+1,0}^B, [\cdot]_{i,k,t,3}^B - [\cdot]_{i,k,t,2}^B = [\cdot]_{i,k,t+1,1}^B - [\cdot]_{i,k,t+1,0}^B \tag{22.j}$$

where $[\cdot]_t^B$ is the spline coefficient vector of $[\cdot](\tau)$ in period $t$. (22.a)-(22.i) are the corresponding representation of (21.a)-(21.i) after BP splines. (22.j) is the first-order continuity constraint for the smooth connection between adjacent periods.

After the reformulation, the CT stochastic power system dispatch model incorporating aggregated HVAC reserve becomes a mixed-integer linear programming problem, which can be tractably solved by commercial solvers.

### *C. Overall Hierarchical Dispatch Framework*

The overall hierarchical dispatch framework proposed in this paper is implemented in the day-ahead dispatch to determine the optimal reserve provision of HVACs, as well as the unit commitment. The implementation procedure is shown in Fig. 4, which includes the following four processes:

***Process 1*** *(Parameter Collection)*: The HVAC aggregators collect the parameters of HVACs that they manage, and then build the base set (14) in preparation for aggregation.

***Process 2*** *(Parallel Aggregation)*: The HVAC aggregators calculate the aggregation models of HVACs via (20) to obtain the optimized value of the affine matrices $\boldsymbol{\Gamma}_{i,k}^{aff}$ and translation vectors $\boldsymbol{\gamma}_{i,k}^{aff}$ in parallel. Then, the aggregators submit the aggregated dispatchable region of HVACs to the power system operator.

***Process 3*** *(Dispatch Distribution)*: The power system operator receives the aggregated dispatchable region of HVACs and centrally calculates the optimal dispatch results via (22). Then, the system operator distributes the power dispatch results to HVAC aggregators.

***Process 4*** *(Parallel Response)*: The HVAC aggregators respond to the power dispatch results in parallel, and allocate the results to the HVACs that they manage.

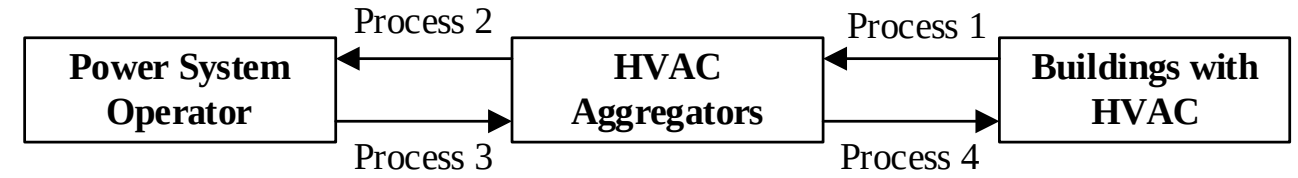

Fig. 4. Diagram of hierarchical dispatch framework.

Two remarks about the framework are given below:

(1) In ***Process 4***, it is no need to solve any disaggregation

optimization problems. By calculating the algebraic equation (23), the aggregator can directly disaggregate the dispatching signals to each HVAC whose power is always feasible. Herein, $[\cdot]^*$ is the optimized value of $[\cdot]$.

$$\boldsymbol{P}_{i,k,s}^{BB,*}=\gamma_{i,k}^{aff,BB}+\boldsymbol{\Gamma}_{i,k}^{aff,BB}\left(\sum_k(\boldsymbol{\Gamma}_{i,k}^{aff,BB})\right)^{-1}\left(\boldsymbol{P}_{i,s}^{app,BB,*}-\sum_k(\gamma_{i,k}^{aff,BB})\right) \tag{23}$$

(2) The above four processes are implemented in sequence and the proposed framework is non-iterative and parallel-enabled. Therefore, it follows the general hierarchical dispatch pattern of real-world power system and has low implementation barriers.

## V. Case Studies

Case studies on the modified 6-bus and IEEE 118-bus test systems are conducted in this section to validate the effectiveness and scalability of the proposed aggregation model. The simulation is on the MATLAB 2021a platform, and uses the commercial solver GUROBI 11.0.0 for calculation. The CPU configuration herein is Intel(R) Core(TM) i7-10875H CPU @ 2.30GHz and the RAM is 16 G.

The 6-bus system is first used to show the advantages of our model, using the data given in [29]. There are three thermal units (G1-G3) and a wind farm with 110MW capacity. The rated total load is 210MW. The dispatch horizon is 24 hours with 1-hour granularity for each period. The load curve, wind power curve, and outdoor temperature curve come from the PJM market and are properly scaled to fit the system. Fig. 5 shows the 5-minute forecast curves and the corresponding DT/CT power and temperature curves in DT/CT dispatch. Ref. [30] is referred to generate and reduce the stochastic scenarios of wind power as shown in Fig. 16 in Appendix D. The penalty coefficients of load shedding and wind curtailment are set as 1000\$/MWh and 200\$/MWh, respectively [31]. The price of reserve deployment for thermal units are assumed to be 1.3 times their highest incremental price [32].

There are 40 HVACs connected at bus 3 within one aggregator. The heterogeneous parameters of HVACs are given in TABLE II. $\delta$ refers to the heterogeneity of HVAC parameters, and is set as 1 unless specified. The risk level $\varepsilon_m$ is set as 1% and the variance $\sigma$ is set as 0.01 in DRCCs. For HVACs, the price of reserve deployment and reserve capacity are set 13\$/MWh and 4\$/MWh, respectively [3].

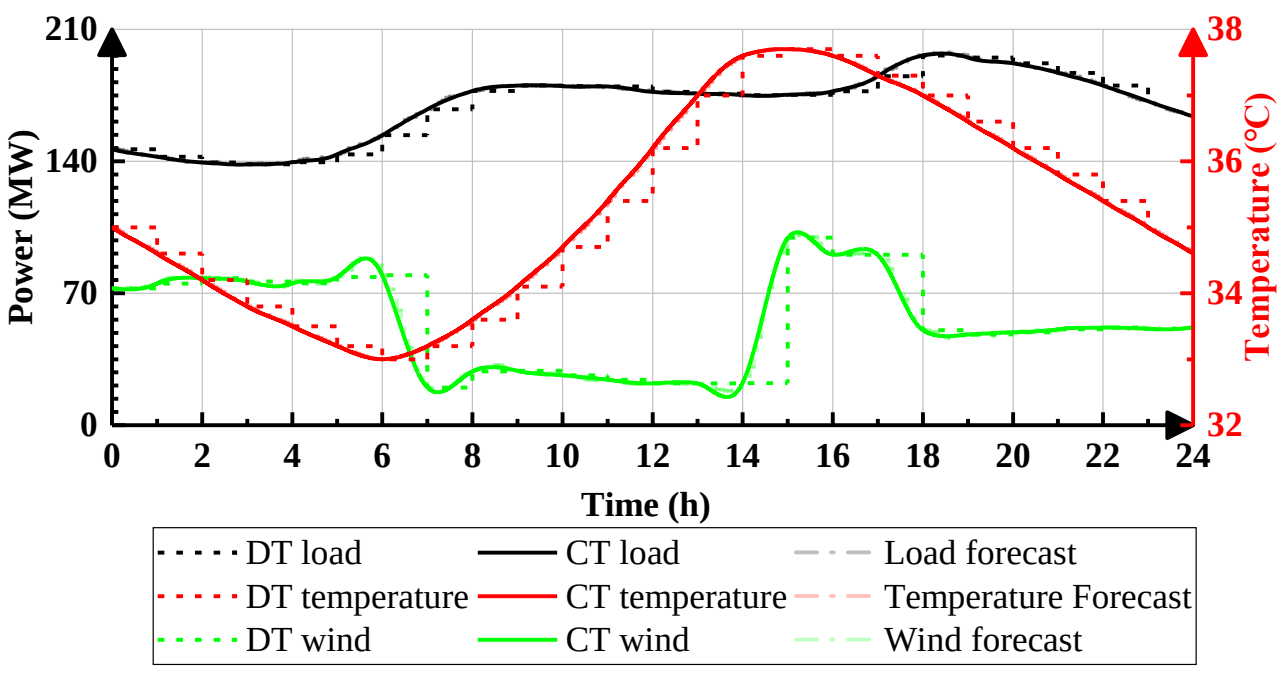

Fig. 5. Total load demand, wind power and outdoor temperature profiles in a typical day.

TABLE II
HVAC Parameters in 6-Bus System

| Item | Value | Item | Value |
|---|---|---|---|
| $H^a$ | $0.25 \pm 0.1\delta$ °C/MW | $H^m$ | $0.5 \pm 0.3\delta$ °C/MW |
| $C^a$ | $2 \pm 1.5\delta$ MWh/°C | $C^m$ | $6 \pm 2\delta$ MWh/°C |
| $T^{set}$ | $21 \pm 1\delta$ °C | $\beta^{set}$ | $1.5 \pm 1\delta$ °C |
| $\mu^h$ | 0.95 | $f^h$ | 0.1 |
| $P^{\min}$ | 0.15 MW | $P^{\max}$ | 0.65 MW |

### *A. Validation of Aggregated Model for HVACs*

With the above settings, we first compare the following cases to validate the aggregation accuracy of our model:

**Case 1**: CT aggregation model using the box-based inner-approximation.

**Case 2**: CT aggregation model using the traditional polytope-based inner-approximation where the scaling factors are same in all dimensions.

**Case 3**: CT aggregation model using the outer-approximation.

**Case 4 (Benchmark)**: CT direct control model without aggregation.

**Case 5 (Proposed)**: CT aggregation model using our proposed affine-approximation.

For polytope-based models, i.e., **Case 2** and **Case 5**, the aggregated HVAC power can be disaggregated directly via (23). For **Case 1** and **Case 3**, each aggregator needs to solve (24) for power disaggregation.

$$\min_{\boldsymbol{P}_{i,k,s,t}^{BB}} \sum_t \left\| \boldsymbol{P}_{i,s,t}^{app,BB,*} - \sum_k \left( \boldsymbol{P}_{i,k,s,t}^{BB} \right) \right\| \tag{24.a}$$

$$\text{s.t.} \quad \boldsymbol{P}_{i,k,s,t}^{BB} \in \mathbb{U}_{i,k} \tag{24.b}$$

For each scenario, the objective function is to minimize the total discrepancy between the sum of HVAC powers and the reference aggregated power. If the discrepancy is not zero, it means that the disaggregation is infeasible.

The results are shown in TABLE III and Fig. 6-10.

Since the aggregation optimization problem (20) of each HVAC can be implemented in parallel and the disaggregation process can also be implemented via calculating the algebraic equation (23) in parallel, the most CPU time used in the proposed framework depends on the time of solving (22). Thus, the CPU time in this paper refers to the time of solving (22).

TABLE III
Dispatch Results of CT Aggregation in 6-Bus System

| Comparison terms | Case1 | Case2 | Case3 | Case4 | Case5 |
|---|---|---|---|---|---|
| First stage energy cost of thermal units (k\$) | 44.16 | 44.16 | 44.16 | 44.16 | 44.16 |
| HVAC upward reserve capacity cost (\$) | 203.30 | 209.61 | 478.81 | 358.18 | 357.56 |
| HVAC downward reserve capacity cost (\$) | 307.88 | 416.33 | 486.13 | 418.30 | 402.84 |
| I: First stage HVAC reserve capacity cost (\$) | 511.18 | 625.94 | 964.94 | 776.48 | 760.37 |

| Expected second stage upward regulation cost of thermal units ($) | 946.43 | 688.86 | 563.04 | 681.43 | 713.96 |
|---|---|---|---|---|---|
| Expected second stage downward regulation cost of thermal units ($) | 1106.2 | 1083.4 | 476.94 | 687.15 | 688.62 |
| Expected second stage HVAC upward regulation cost ($) | 315.28 | 328.48 | 558.48 | 520.18 | 519.48 |
| Expected second stage HVAC downward regulation cost ($) | 442.04 | 565.01 | 562.57 | 568.56 | 553.03 |
| Expected second stage wind curtailment cost ($) | 74.52 | 38.45 | 0 | 0 | 0 |
| II: Expected second stage cost (k$) | 2.88 | 2.71 | 2.16 | 2.46 | 2.48 |
| I+II (k$) | 3.40 | 3.34 | 3.13 | 3.23 | 3.24 |
| Total HVAC reserve capacity (MWh) | 127.80 | 156.48 | 241.24 | 194.12 | 190.10 |
| Infeasibility rate | 0 | 0 | 28% | 0 | 0 |
| Number of continuous variables | 33,586 | 33,586 | 33,586 | 156,046 | 33,586 |
| Number of binary variables | 504 | 504 | 504 | 504 | 504 |
| CPU time (s) | 1.18 | 1.39 | 1.37 | 64.14 | 1.75 |

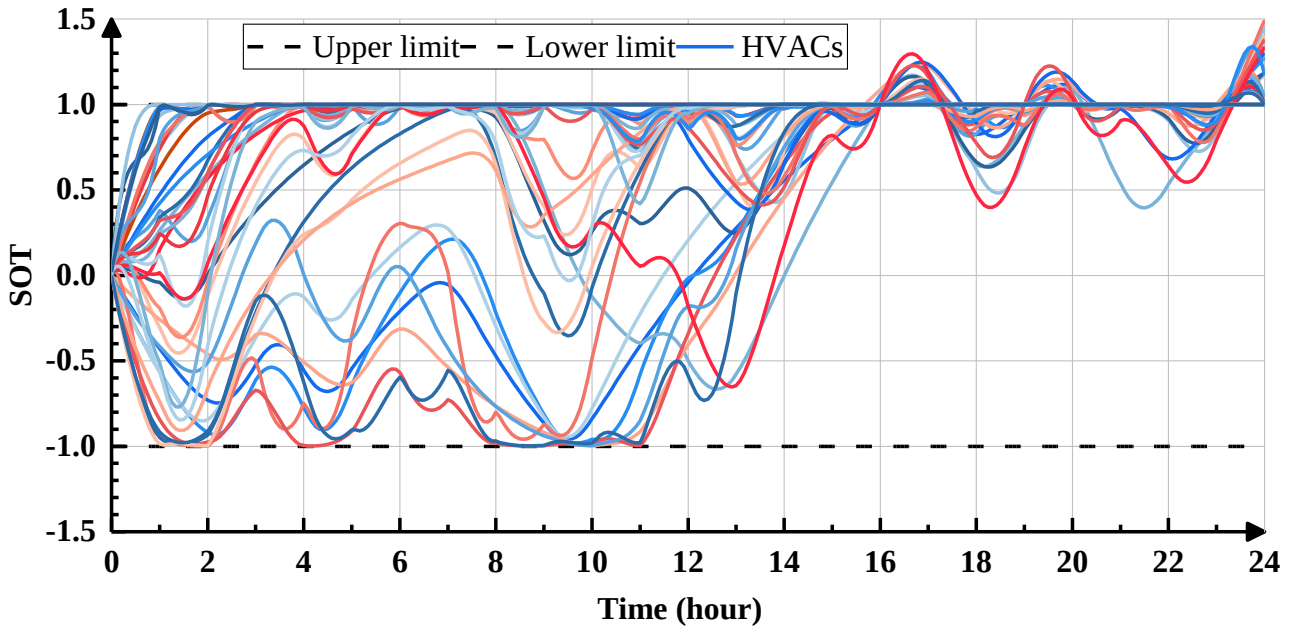

Fig. 6. Disaggregation result of HVAC indoor temperature trajectories in one scenario of **Case 3.** ($SOT$, i.e., the state of indoor temperature [33], is defined as $(T^{a}-T^{set})/\beta^{set}$ for analysis and bounded between -1 and 1 according to (1.d))

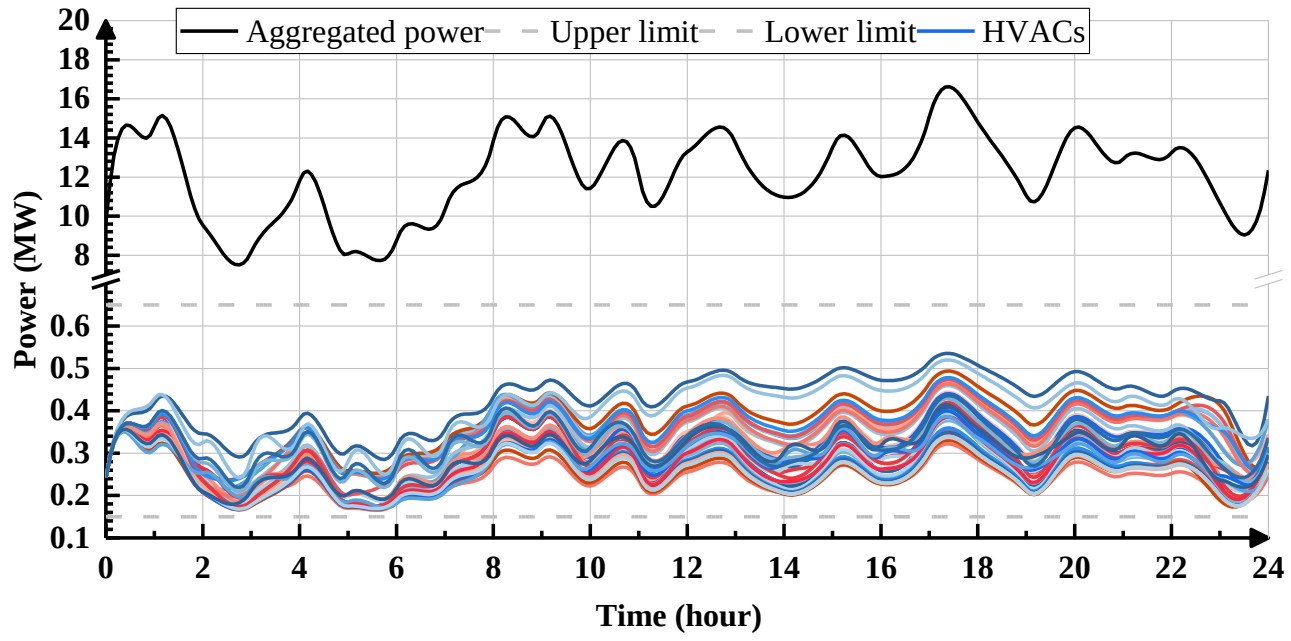

Fig. 7. Disaggregation result of HVAC power trajectories in one scenario of **Case 5.**

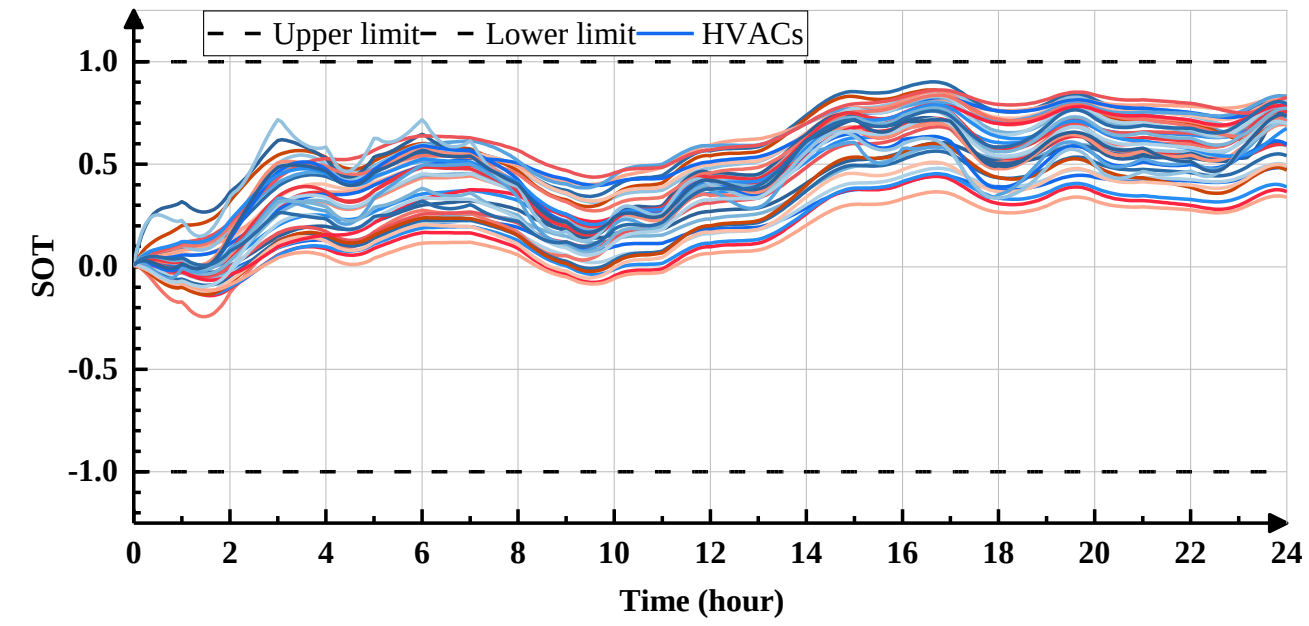

Fig. 8. Disaggregation result of HVAC indoor temperature trajectories in one scenario of **Case 5**.

**Case 4** shows the most CPU time since individual HVAC model is comprehensively considered to guarantee the optimality. However, it may not be applicable in practical HVAC scheduling due to its heavy computational burden. Thus, we only use **Case 4** as **Benchmark** to evaluate the aggregation accuracy of different aggregation models.

**Case 3** shows the highest HVAC reserve capacity among aggregation models in TABLE III. Since the reserve of HVACs is more cost-effective than that of thermal units, **Case 3** shows the lowest I+II cost (even lower than **Benchmark**). Indeed, the outer approximation model in **Case 3** ignores the parameter heterogeneity of HVACs and directly sums the constraint coefficients of (1), obtaining a larger aggregated feasible region than the exact one. Hence, it cannot guarantee disaggregation feasibility and has a 28% infeasible rate during power disaggregation. We show the disaggregation results of temperature trajectories in one scenario of **Case 3** in Fig. 6 where the thermal discomfort periods of all users are 5 hours.

On the other hand, our model in **Case 5**, although being conservative, enjoys 100% disaggregation feasibility which is guaranteed by the containment constraint (17). We show the disaggregation results of power and temperature trajectories in our model in Fig. 7-8 where the operational constraints of HVACs are strictly respected. The above results demonstrate the disaggregation feasibility of our model.

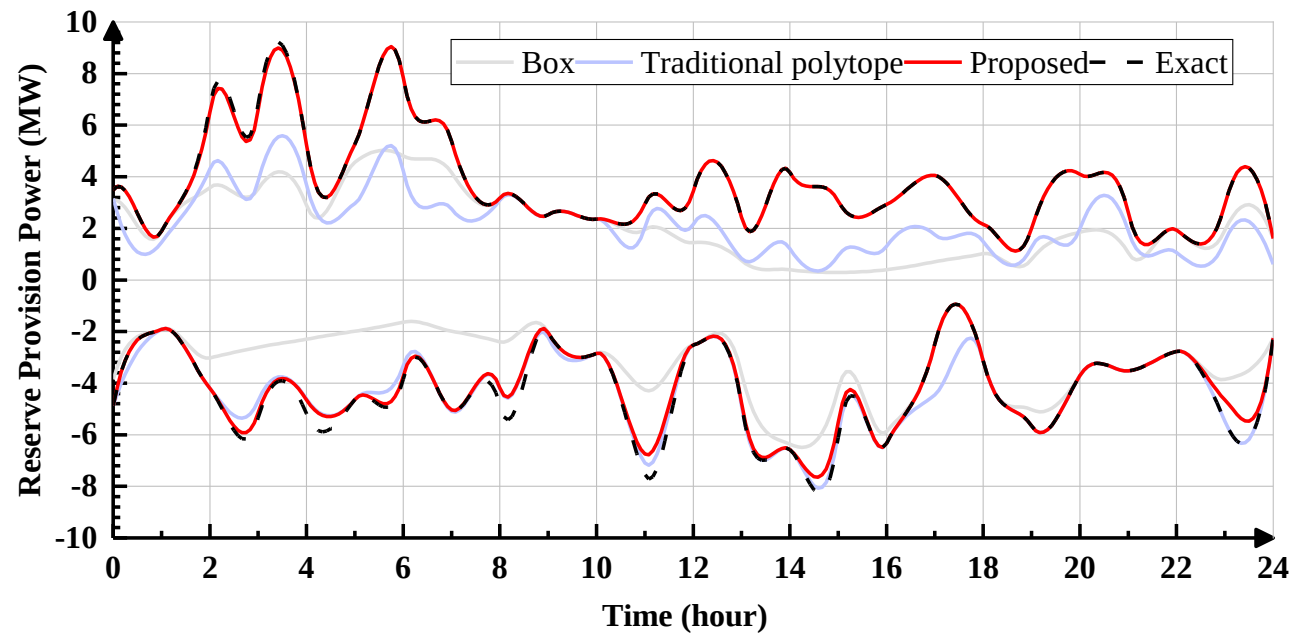

Fig. 9. The reserve capacity bounds of aggregated HVACs.

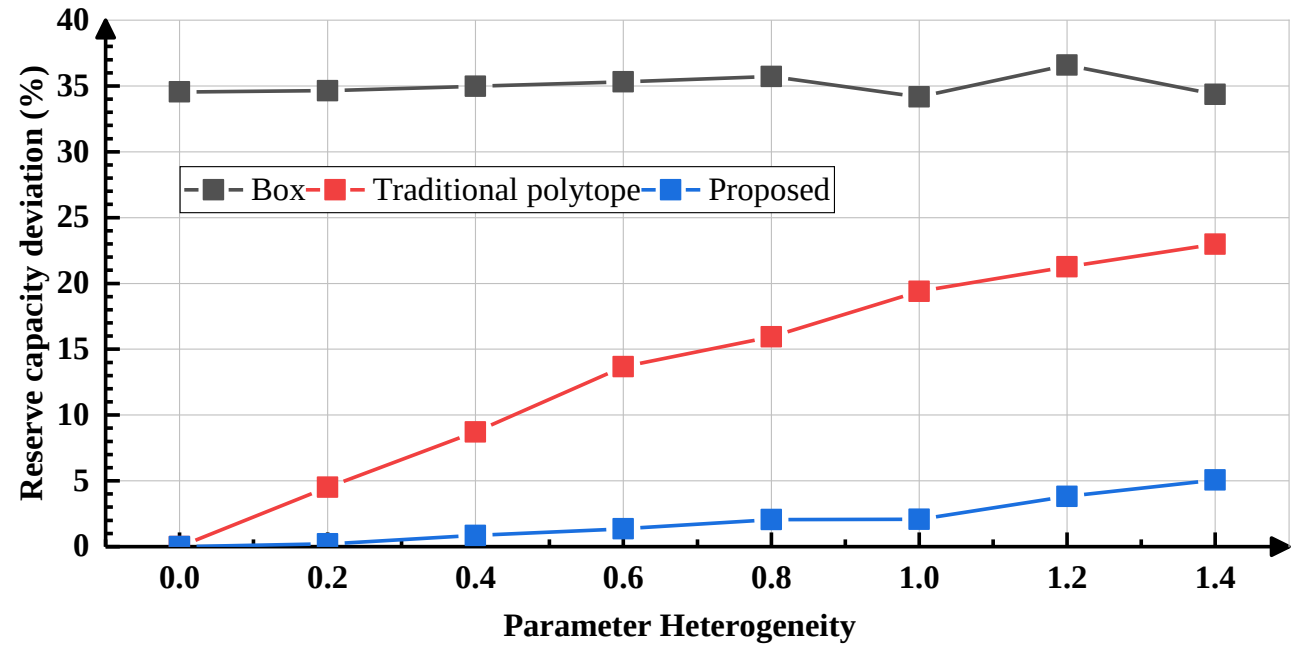

Fig. 10. Sensitivity analysis of parameter heterogeneity $\delta$. (The deviation is defined as the absolute relative deviation of the reserve capacity between other cases and **Benchmark**)

In TABLE III, compared with other inner-approximation models (**Case 1** and **Case 2**), our model identifies more cost-effective HVAC reserve capacity in the first stage and there is no wind curtailment in the second stage which occurs in **Case 1** and **Case 2**. Hence, our model enjoys the lower total cost and obtains the closer solution to **Benchmark**.

The reserve capacity bounds of different models are given in Fig. 9. **Case 1** and **Case 2** preserve the structure of base set and sacrifice the great feasible region of HVACs, thereby having quite limited adjustable range of HVAC capacity bounds. In contrast, our model can adjust the structure of base set by the proposed affine transformation to adapt to the high-dimensional feasible region of HVACs, thereby retaining the most flexibility and having larger adjustable capacity bounds.

We further conduct the sensitivity analysis of parameter heterogeneity in Fig. 10. The performance of **Case 1** is the worst compared to other models due to the simplicity of base set. For $\sigma = 0$ (homogeneous parameters for HVACs), both **Case 2** and our model identify the optimal HVAC reserve capacity. As the degree of heterogeneity increases, **Case 2** loses more flexibility due to its poor compatibility. Instead, our model has great geometric adaptability and thereby has the lowest deviation. Although our model enjoys the best performance, we also observe that the deviation of our model reaches to over 5% when the parameter heterogeneity is greater than 1.4. We note that this is a necessary trade-off for guaranteeing computational tractability.

The CPU time of our model is slightly more than other models due to its general transformed structure of base set, which is paid for the high aggregation accuracy. Also, in TABLE III, compared to **Benchmark**, our aggregation model can sharply reduce the scale of the dispatch problem while obtaining the near-optimal solution.

The above results show that our aggregation model can effectively hedge against the heterogeneity in high-dimensional feasible region with high aggregation accuracy and relatively low computational complexity.

*B. Advantage of CT Aggregation*

To evaluate the advantage of our CT aggregation, we further compare the following cases:

**Case 5 (Proposed)**: CT aggregation model using our proposed affine approximation.

**Case 5a**: Traditional DT aggregation model using our proposed affine approximation.

Note that the proposed two-stage stochastic optimization is conducted day-ahead to determine commitment decisions and HVAC reserve capacity. After these day-ahead decisions have been decided, we conduct a real-time economic dispatch simulation with 5-minute granularity (using data in Fig. 5 and Fig. 16) to evaluate the performance of day-ahead decisions, where all commitment decisions and reserve capacity are fixed to the optimal values from the solution of two-stage stochastic optimization. Since the economic dispatch has no binary variables, it is a simple linear programming problem. The results are reported in TABLE IV and Fig. 11-12.

In TABLE IV, more HVAC reserve capacity and thermal unit generation are prepared in the first stage of our model Thus, the first stage total cost of our model is slightly higher than that of **Case 5a**. However, due to the ignorance of intra-hour HVAC flexibility, the latent HVAC reserve capacity is not fully exploited in **Case 5a**. Thus, more cost-expensive upward/downward regulation of thermal units is required in real-time operation of **Case 5a** to make up for the HVAC reserve shortage. Also, there are some wind curtailment and load shedding in **Case 5a** due to the flexibility shortage, which further worsen its economy. As a result, the total cost of **Case 5a** is higher than our model.

More specifically, from Fig. 11-12, we observe that load shedding occurs at 13:20-13:50 in **Case 5a** due to the insufficient downward regulation capacity of HVAC reserve. Also, there is some wind curtailment at 1:40-2:00, 5:20-5:50, and 14:30-15:05 in **Case 5a** since the HVACs and thermal units cannot follow the fast real-time net load variation. Especially at 14:30-15:05, the HVAC upward regulation capacity of HVACs is quite limited, leading to a large amount of wind curtailment.

The CPU time is also provided in TABLE IV. Compared to the traditional DT aggregation, more CPU time is required in our model due to the existences of more variables, which is paid for the intra-hour flexibility utilization.

The above results show that our model can identify the potential intra-hour aggregated HVAC flexibility, contributing to the more secure and economic operation for power system dispatch.

TABLE IV
DISPATCH RESULTS OF CT AND DT AGGREGATION CASES

| Comparison terms | **Case 5**: proposed CT aggregation | **Case 5a**: traditional DT aggregation |
|---|---|---|
| First stage energy cost of thermal units (k$) | 44.16 | 42.87 |
| HVAC upward reserve capacity cost ($) | 357.56 | 296.58 |
| HVAC downward reserve capacity cost ($) | 402.84 | 302.82 |
| First stage total cost (k$) | 44.92 | 43.47 |
| Expected real-time upward regulation cost of thermal units ($) | 405.70 | 2157.20 |
| Expected real-time downward regulation cost of thermal units | 366.19 | 1621.29 |
| ($) | | |
| Expected real-time HVAC upward regulation cost ($) | 441.14 | 352.98 |
| Expected real-time HVAC downward regulation cost ($) | 482.86 | 389.26 |
| Expected real-time load shedding cost ($) | 0 | 117.70 |
| Expected real-time wind curtailment cost ($) | 0 | 105.78 |
| Expected real-time cost ($) | 1695.9 | 4744.21 |
| Total cost (k$) | 46.62 | 48.21 |
| Day-ahead total HVAC reserve capacity (MWh) | 190.10 | 149.85 |
| Number of continuous variables | 33,586 | 11,654 |
| Number of binary variables | 504 | 216 |
| Day-ahead CPU time (s) | 1.75 | 0.83 |

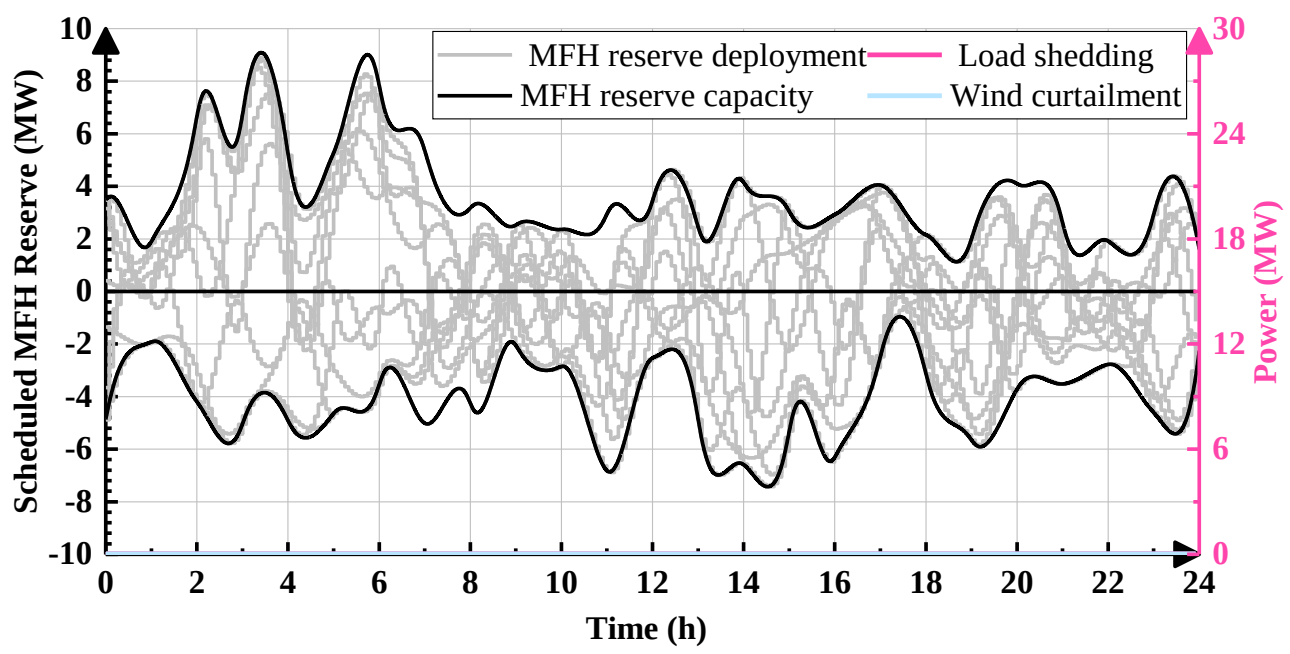

Fig. 11. Scheduled CT reserve for HVACs.

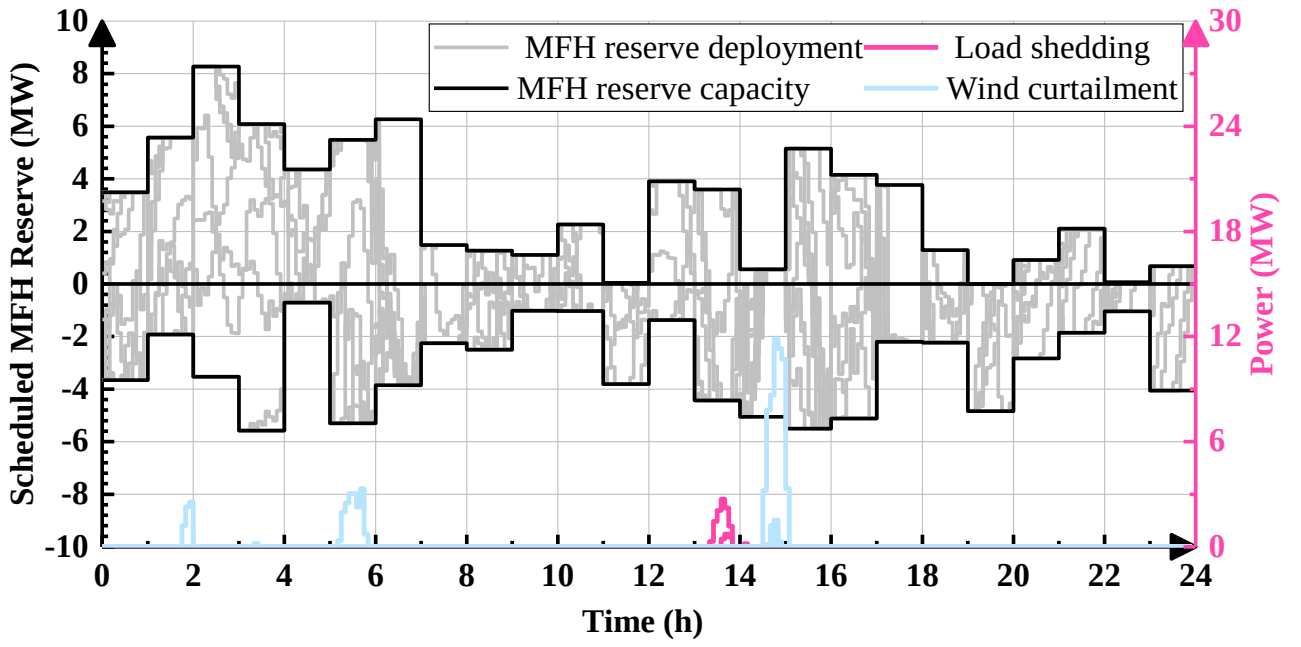

Fig. 12. Scheduled DT reserve for HVACs.

### *C. Impact of Outdoor Temperature Uncertainty*

To evaluate the impact of outdoor temperature uncertainty, we compare the following cases:

**Case 5 (Proposed)**: CT aggregation model considering the temperature uncertainty.

**Case 5b**: CT aggregation model ignoring the temperature uncertainty (the variance is set as zero).

The results are shown in TABLE V and Fig. 13-15.

TABLE V
COMPARISONS OF UNCERTAINTY IMPACT

| Comparison terms | **Case 5**: considering uncertainty | **Case 5b**: ignoring uncertainty |
|---|---|---|
| I: First stage reserve capacity cost ($) | 760.37 | 868.872 |
| II: Second stage cost ($) | 2475.09 | 2306.69 |
| I+II ($) | 3235.09 | 3175.562 |
| Day-ahead total HVAC reserve capacity (MWh) | 190.10 | 217.218 |

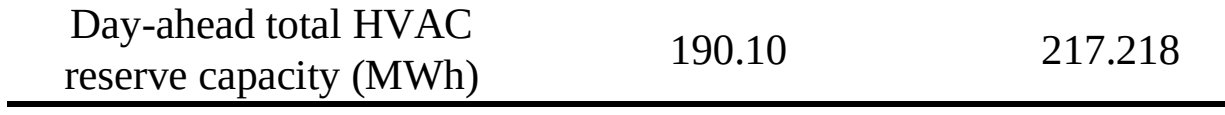

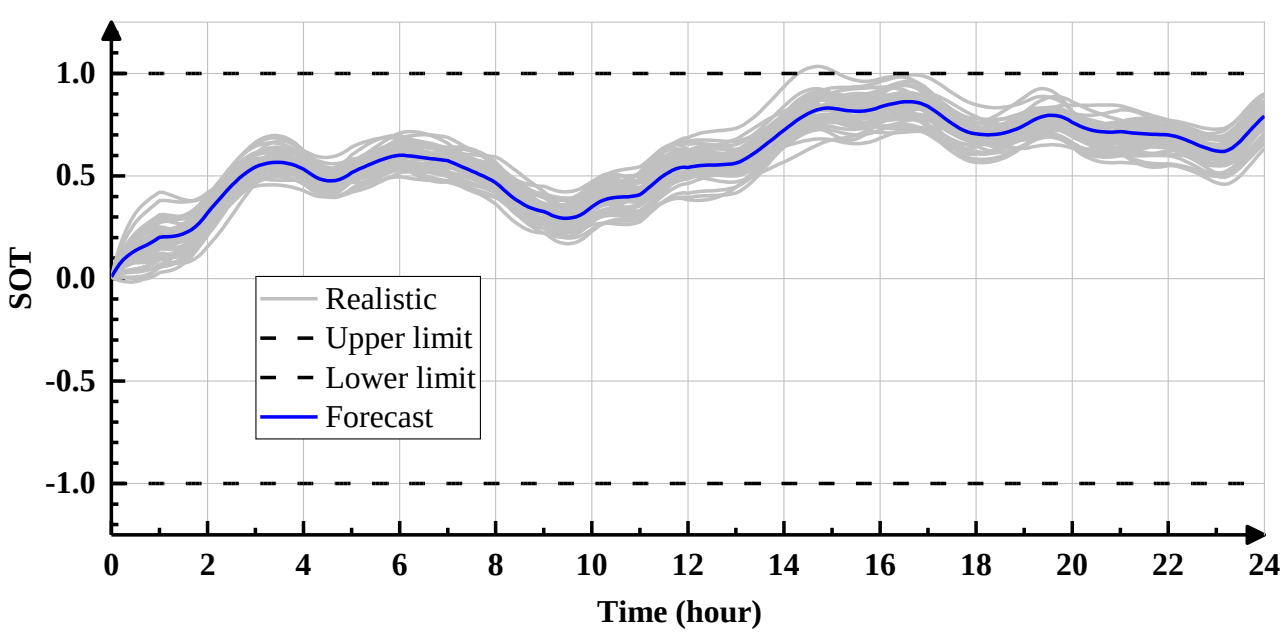

Fig. 13. HVAC's indoor temperature in **Case 5**.

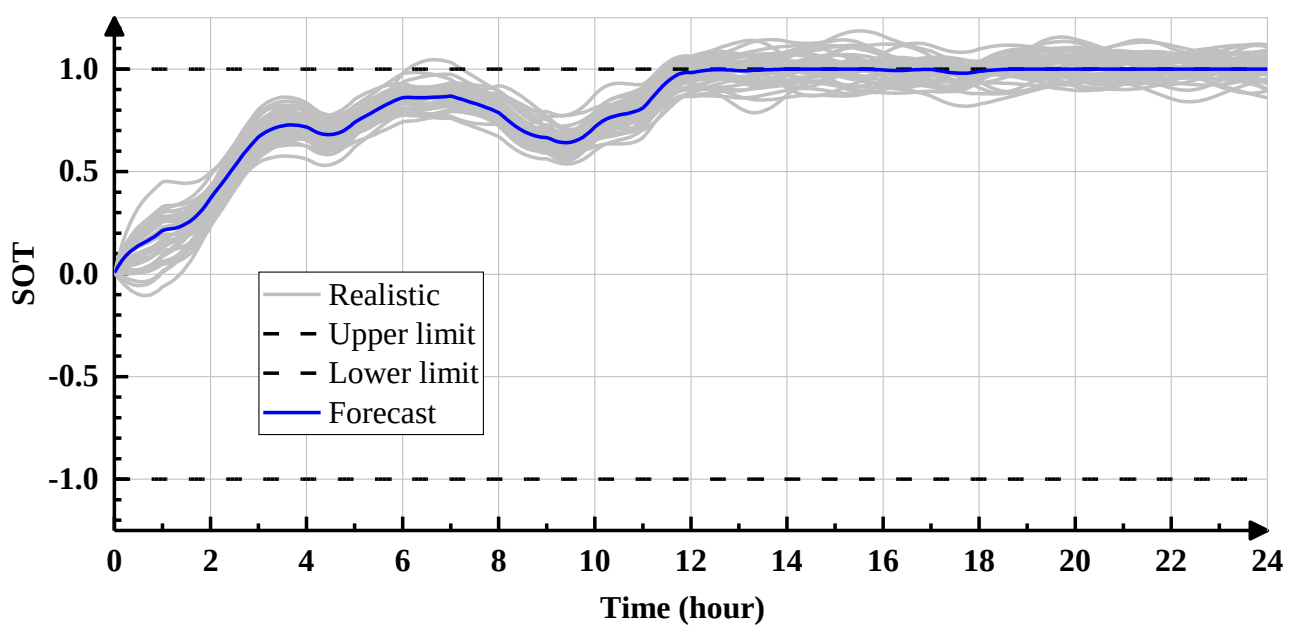

Fig. 14. HVAC's indoor temperature in **Case 5b**.

From TABLE V, we observe that more cost-effective HVAC reserve capacity is given when the outdoor temperature uncertainty is ignored, which means larger HVAC flexibility is identified to reduce the total I+II cost. However, the dispatch results of **Case 5b** may not be applicable in practice when the uncertain outdoor temperature is considered.

To further test the feasibility of HVAC dispatch results, we select the scheduling plan of an HVAC and randomly generate 1000 actual scenarios of outdoor temperature via Monte Carlo sampling. The testing results of indoor temperature are given in Fig. 13-14 where the solid blue line denotes the forecasted indoor temperature of the HVAC under the forecasted scenario, and gray lines denote temperature variation under realistic outdoor temperature scenarios. We observe that the forecast indoor temperature will "hit the upper bound" in **Case 5b**. Thus, the violation of indoor temperature limit occurs when the forecasting error is considered. In contrast, a certain margin to the indoor temperature bound is reserved in our model, effectively mitigating the forecasting error.

We further vary the risk level $\varepsilon^m$ in our model to evaluate its impact on the I+II cost as shown in Fig. 15. We observe that more positive the risk attitude is, more aggregation feasible region (proportional to the total trace of $\boldsymbol{\Gamma}_{i,k}^{aff,BB}$) the system obtains, which reduces more system costs.

The above results show that our aggregation model can efficiently handle the impact of outdoor temperature uncertainty for reliable aggregation. In addition, our model with the controllable parameter, i.e., $\varepsilon^m$, provides a regulation measure to balance economy and risk attitude in real applications.

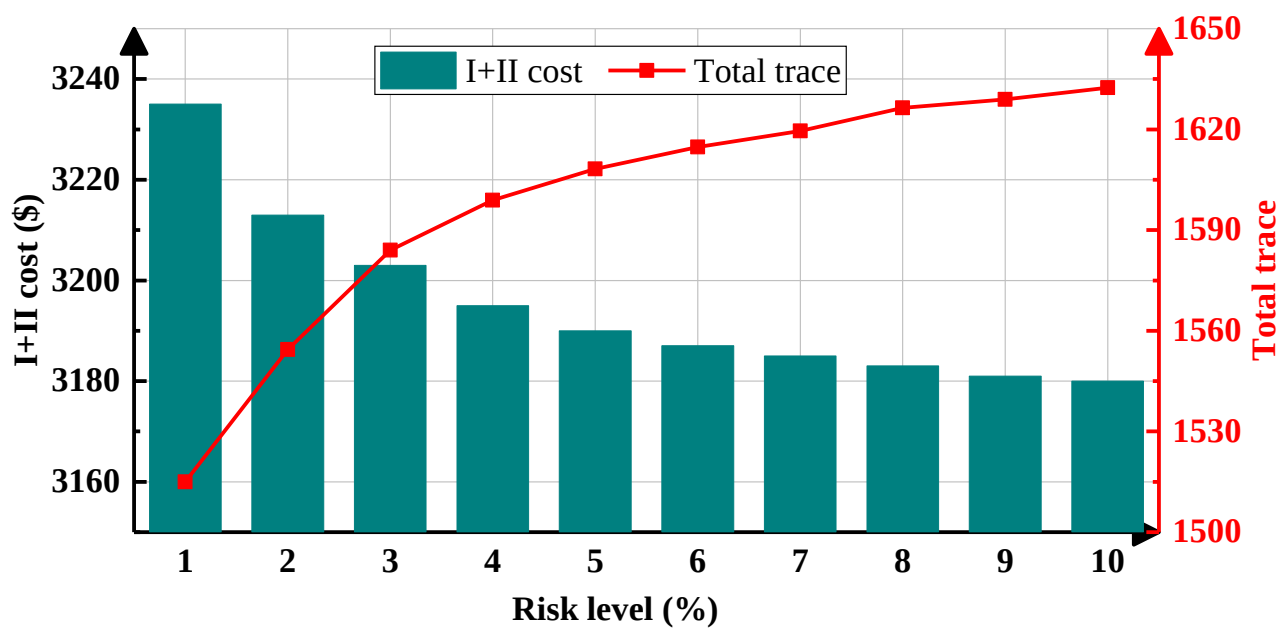

Fig. 15. DRCC sensitivity analysis.

## D. Scalable Test

We further verify the scalability and the reasonable computation complexity of our model using the IEEE 118-bus system, which includes 54 thermal units and a 4405 MW of maximum total load. The parameters of thermal units can be referred to [34]. The wind capacity is scaled up to 1600MW. The forecast curves of load, wind power, and outdoor temperature are similar to Fig. 5 and Fig. 16, and are also scaled up to fit the 118-bus system. The penalty coefficients of load shedding and wind curtailment are set as the same as the 6-bus system. There are 1200 HVACs averagely distributed in 3 groups whose parameters are from TABLE VI. The risk level, the variance, and the cost coefficients are set as the same as the 6-bus system.

TABLE VI
HVAC PARAMETERS IN 118-BUS SYSTEM

| Item | Group 1 | Group 2 | Group 3 |
|---|---|---|---|
| $H^a$ (°C/MW) | 0.1-0.25 | 0.1-0.15 | 0.25-0.4 |
| $H^m$ (°C/MW) | 0.5-2 | 0.3-1 | 0.1-0.2 |
| $C^a$ (MWh/°C) | 1-3 | 2.5-5.5 | 3-7 |
| $C^m$ (MWh/°C) | 3-6 | 3-5 | 5-8 |
| $T^{set}$ (°C) | 17-21 | 19-23 | 20-22 |
| $\beta^{set}$ (°C) | 1-3 | 1-2 | 2-3 |
| $\mu^h$ | 0.9-0.95 | 0.92-0.94 | 0.92-0.95 |
| $f^h$ | 0.08-0.1 | 0.05-0.1 | 0.08-0.12 |
| $P^{\min}$ (MW) | 0.05-0.1 | 0.03-0.09 | 0.05-0.25 |
| $P^{\max}$ (MW) | 0.6-0.7 | 0.4-0.6 | 0.75-0.85 |

The results are reported in TABLE VII. We observe that **Benchmark** is no longer able to calculate the optimal solution within one day due to its unbearable computational burden, i.e., the massive variables brought by HVACs. In contrast, our model, with the much smaller scale of optimization problem, can be calculated in a reasonable CPU time. Also, our model can identify more flexibility of HVACs compared to the other two inner-approximation aggregation models, since the total HVAC reserve capacity identified in **Case 1** and **Case 2** accounts for only 33.6% and 80.0% of our **Case 5**. Hence, our model can achieve the lowest expected wind curtailment cost and obtain the best economy. The above results demonstrate the scalability of our aggregation model in larger-sized systems.

TABLE VII
SCALABILITY TEST RESULTS IN 118-BUS SYSTEM

| Comparison terms | **Case 1** | **Case 2** | **Case 4** | **Case 5** |
|---|---|---|---|---|
| First stage energy cost of thermal units (k$) | 2122.7 | 2121.0 | / | 2121.3 |
| HVAC reserve capacity cost (k$) | 13.76 | 32.7 | / | 40.9 |
| Expected regulation cost of thermal units (k$) | 446.7 | 262.0 | / | 172.7 |
| Expected HVAC reserve deployment cost (k$) | 19.1 | 35.6 | / | 39.7 |
| Expected wind curtailment cost (k$) | 19.7 | 15.3 | / | 6.0 |
| Total cost (k$) | 2622.0 | 2466.7 | / | 2380.6 |
| Total HVAC reserve capacity (MWh) | 3439.7 | 8177.2 | / | 10225.0 |
| Number of continuous variables | 519,742 | 519,742 | 4,225,512 | 519,742 |
| Number of binary variables | 9,072 | 9,072 | 9,072 | 9,072 |
| CPU time | 418.5 s | 515.4 s | >1 day | 526.3 s |

# VI. CONCLUSIONS

In this paper, we propose the CT aggregation of massive flexible HVAC loads considering uncertainty for reserve provision in power system dispatch. We first develop the novel CT aggregation model of HVACs considering the heterogeneity in high-dimensional feasible region and the impact of outdoor temperature uncertainty. Then, we propose a cascade of tailored reformulation techniques to recast the CT aggregation model as linear programming. Further, we propose a customized hierarchical dispatch framework, incorporating the proposed CT aggregation model for the efficient utilization of massive HVACs to cope with the renewable uncertainty.

Extensive case studies based on modified 6-bus and IEEE 118-bus systems demonstrate effectiveness and scalability of our CT aggregation model in 1) hedging against the heterogeneity in high-dimensional feasible region for high aggregation accuracy; 2) identifying the intra-hour flexibility of HVACs for secure and economic operation in power system dispatch; 3) dealing with the impact of outdoor temperature uncertainty for reliable aggregation and balancing economy and risk attitude in real applications.

For future research, an intriguing avenue involves discussing more measures to address the great parameter heterogeneity in aggregation for more accurate aggregation. Additionally, exploring the common aggregation model under multiple source uncertainties will be another extension to this work.

# APPENDIX

## A. Proof of Inner-approximation

Given a base set as shown in (4.a), the affine-approximation aggregated set $\mathbb{U}_i^{app}$ can be expressed in (25):

$$\begin{aligned}\mathbb{U}_i^{app} &= \left\{ \boldsymbol{P}_i^{app} = \sum_k (\boldsymbol{\Gamma}_{i,k}^{aff})\boldsymbol{P}_i^{base} + \sum_k (\gamma_{i,k}^{aff}), \boldsymbol{P}_i^{base} \in \mathbb{U}_i^{base} \right\} \\ &= \left\{ \boldsymbol{P}_i^{app} = \sum_k (\boldsymbol{\Gamma}_{i,k}^{aff}\boldsymbol{P}_i^{base} + \gamma_{i,k}^{aff}), \boldsymbol{P}_i^{base} \in \mathbb{U}_i^{base} \right\}\end{aligned} \tag{25}$$

Considering that $\boldsymbol{\Gamma}_{i,k}^{aff}\boldsymbol{P}_i^{base} + \boldsymbol{\gamma}_{i,k}^{aff} \subseteq \mathbb{U}_{i,k}^{aff}, \forall k$, we can

further obtain $\mathbb{U}_i^{app} = \sum_k (\boldsymbol{\Gamma}_{i,k}^{aff} \boldsymbol{P}_i^{base} + \boldsymbol{\gamma}_{i,k}^{aff}) \subseteq \biguplus_k \mathbb{U}_{i,k}^{aff}$.

Similarly, considering that $\mathbb{U}_{i,k}^{aff} \subseteq \mathbb{U}_{i,k}, \forall k$ from (7.b), we can obtain $\biguplus_k \mathbb{U}_{i,k}^{aff} \subseteq \biguplus_k \mathbb{U}_{i,k} = \mathbb{U}_i^{agg}$.

*B. Proof of Containment Constraint Reformulation*

Given a base set $\mathbb{U}_i^{base} = \{\boldsymbol{H}_i^{base,BB} \boldsymbol{P}_{i,k}^{BB} \le \boldsymbol{h}_i^{base,BB}\}$, the affine-approximation aggregated set $\mathbb{U}_{i,k}^{aff}$ can be expressed as:

$$\mathbb{U}_{i,k}^{aff} = \{\boldsymbol{P}_{i,k}^{aff,BB} = \boldsymbol{\Gamma}_{i,k}^{aff,BB} \boldsymbol{P}_{i,k}^{BB} + \boldsymbol{\gamma}_{i,k}^{aff,BB}, \forall \boldsymbol{H}_i^{base,BB} \boldsymbol{P}_{i,k}^{BB} \le \boldsymbol{h}_i^{base,BB}\} \tag{26.a}$$

On the other hand, it can be expressed in the form of H-representation of a polytope as follows:

$$\mathbb{U}_{i,k}^{aff} = \{\boldsymbol{H}_{i,k}^{aff,BB} \boldsymbol{P}_{i,k}^{aff,BB} \le \boldsymbol{h}_{i,k}^{aff,BB}\} \tag{26.b}$$

where $\boldsymbol{H}_{i,k}^{aff,BB}$ and $\boldsymbol{h}_{i,k}^{aff,BB}$ are coefficient matrices.

Combining (26.a) and (26.b), we obtain (26.c) as follows:

$$\begin{cases} \boldsymbol{H}_{i,k}^{aff,BB} \boldsymbol{\Gamma}_{i,k}^{aff,BB} = \boldsymbol{H}_i^{base,BB} \\ \boldsymbol{h}_{i,k}^{aff,BB} = \boldsymbol{h}_i^{base,BB} + \boldsymbol{H}_{i,k}^{aff,BB} \boldsymbol{\gamma}_{i,k}^{aff,BB} \end{cases} \tag{26.c}$$

Meanwhile, according to [28], the containment relationship $\mathbb{U}_{i,k}^{aff} \subseteq \mathbb{U}_{i,k}$ can be expressed as follows:

$$\begin{cases} \boldsymbol{\Lambda}_{i,k} \ge 0 \\ \boldsymbol{\Lambda}_{i,k} \boldsymbol{H}_{i,k}^{aff,BB} = \boldsymbol{H}_{i,k}^{BB} \\ \boldsymbol{\Lambda}_{i,k} \boldsymbol{h}_{i,k}^{aff,BB} \le \boldsymbol{h}_{i,k}^{BB} \end{cases} \tag{26.d}$$

where $\boldsymbol{\Lambda}_{i,k}$ is the auxiliary variable matrix.

Combining (26.c) and (26.d), we can obtain (17.a)-(17.c).

*C. Proof of the Reformulation of DRCCs*

We follow the widely used the Bonferroni approximation [24] which set the violation probability $\varepsilon_m$ of the constraint at row $m$ to a fixed value as below:

$$\begin{cases} \inf_{f(\boldsymbol{h}_{i,k}^{BB}) \in \mathcal{D}} \mathbb{P}_{\boldsymbol{h}_{i,k}} \begin{Bmatrix} \boldsymbol{\Lambda}_{i,k}(m,:) \boldsymbol{h}_i^{base,BB} \\ + \boldsymbol{H}_{i,k}^{BB}(m,:) \boldsymbol{\gamma}_{i,k}^{aff,BB} \le \boldsymbol{h}_{i,k}^{BB}(m) \end{Bmatrix} \ge 1 - \varepsilon_m \\ \sum_m \varepsilon_m \le \varepsilon, \varepsilon_m \ge 0 \end{cases} \tag{27.a}$$

Thus, for a row $m$ in $\boldsymbol{h}_{i,k}^{BB}$ that does not contain an uncertain parameter, (18) is equivalent to:

$$\boldsymbol{\Lambda}_{i,k}(m,:) \boldsymbol{h}_i^{base,BB} + \boldsymbol{H}_{i,k}^{BB}(m,:) \boldsymbol{\gamma}_{i,k}^{aff,BB} \le \boldsymbol{h}_{i,k}^{BB}(m), m \in M_d \tag{27.b}$$

For rows containing uncertain parameter in $\boldsymbol{h}_{i,k}^{BB}$, according to the Theorem in [24], (18) can be expressed equivalently as:

$$\begin{aligned} &\boldsymbol{\Lambda}_{i,k}(m,:) \boldsymbol{h}_i^{base,BB} + \boldsymbol{H}_{i,k}^{BB}(m,:) \boldsymbol{\gamma}_{i,k}^{aff,BB} \\ &\le \boldsymbol{\mu}_{i,k}(m) - \boldsymbol{\sigma}_{i,k}(m) \sqrt{\frac{1-\varepsilon_m}{\varepsilon_m}}, m \in M_u \end{aligned} \tag{27.c}$$

where $\varepsilon_m = \varepsilon / L_{\text{dim}}$. $L_{\text{dim}}$ is the number of elements contained in $M_u$. Then, the proof is completed.

*D. Stochastic Scenarios of Wind Power in 6-bus System*

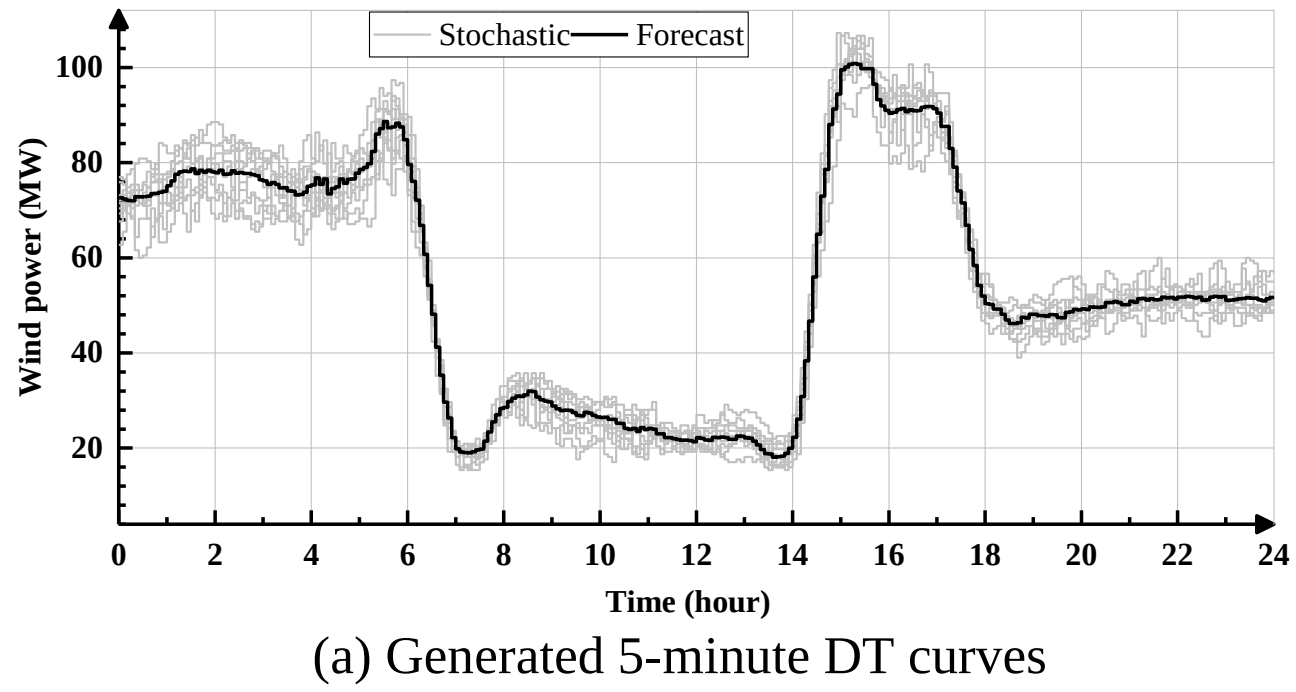

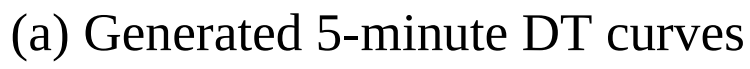

(a) Generated 5-minute DT curves

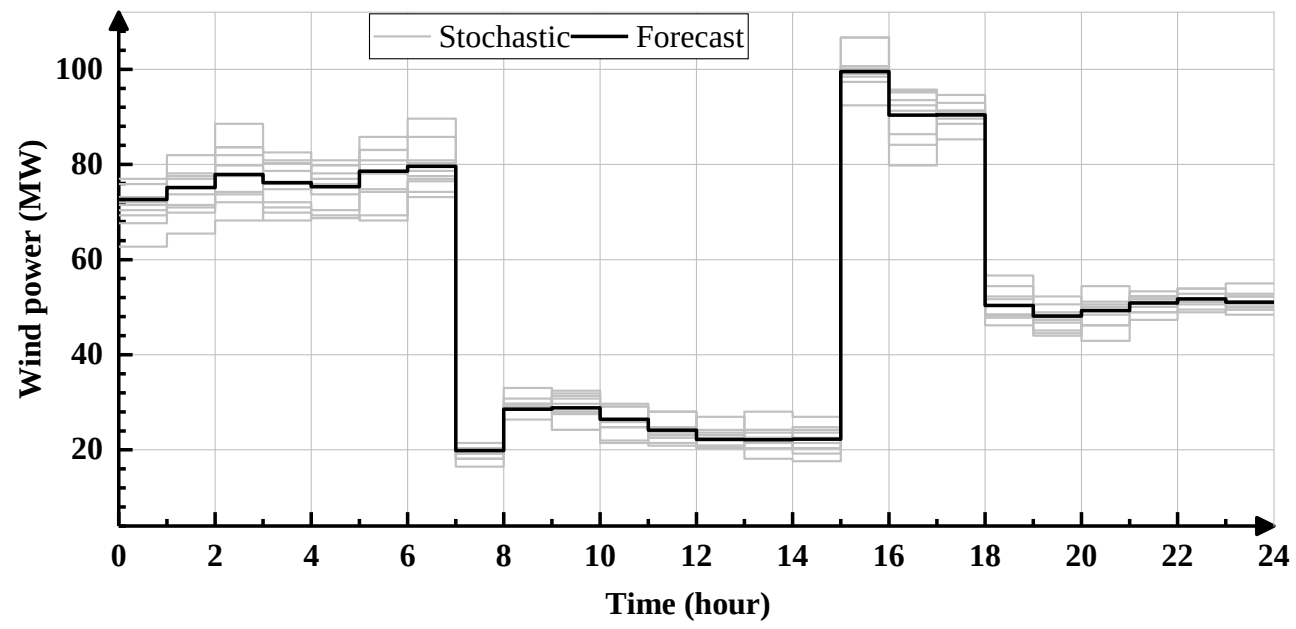

(b) Wind power curves in DT dispatch

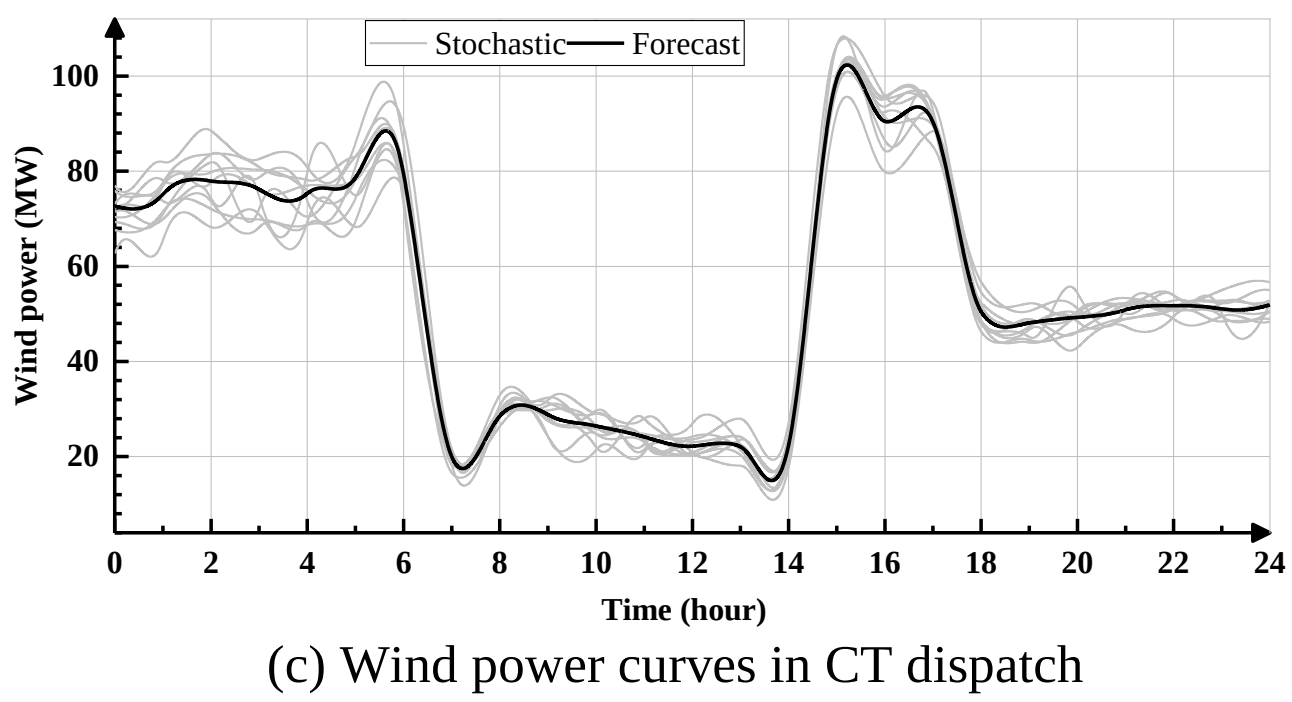

(c) Wind power curves in CT dispatch

Fig. 16. Stochastic scenarios of wind power.

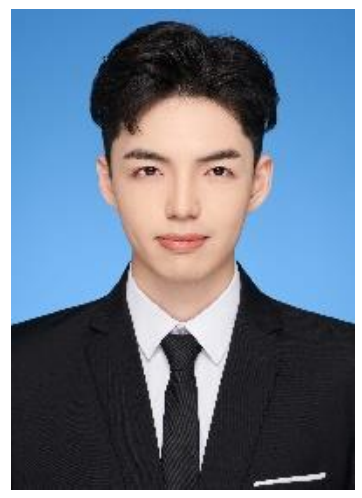

**Jingguan Liu** (Student Member, IEEE) received the B.S. degree in 2022 in electrical engineering from the Huazhong University of Science and Technology, Wuhan, China, where he is currently working toward the Ph.D. degree in electrical engineering.

His current research interests include aggregation of flexible loads, flexible operation of power systems, and continuous-time scheduling.

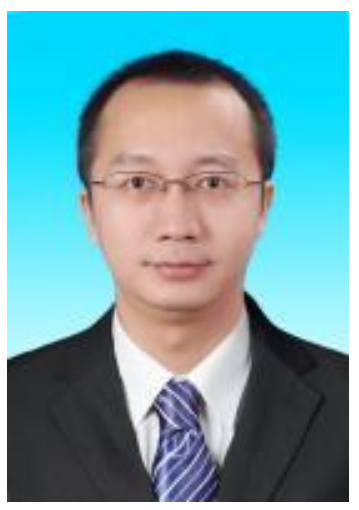

**Xiaomeng Ai** (Member, IEEE) received the B.Eng. degree in mathematics and applied mathematics and the Ph.D. degree in electrical engineering from the Huazhong University of Science and Technology (HUST), Wuhan, China, in 2008 and 2014 respectively.

He is currently a Professor with the School of Electrical and Electronics Engineering, HUST. His research interests include robust optimization theory in power system, renewable energy integration, and integrated energy market.

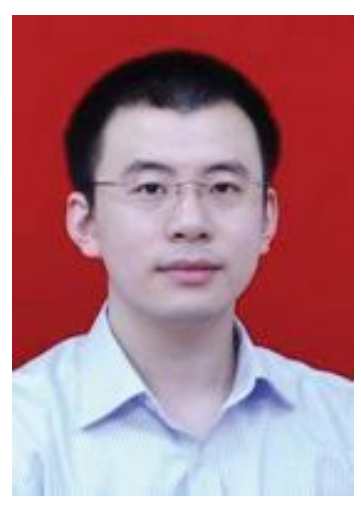

**Jiakun Fang** (Senior Member, IEEE) received the B.Sc. and Ph.D. degrees from the Huazhong University of Science and Technology (HUST), Wuhan, China, in 2007 and 2012, respectively.

From 2012 to 2019, he was with the Department of Energy Technology, Aalborg University, Aalborg, Denmark. He is currently a Professor with the School of Electrical and Electronics Engineering, Huazhong University of Science and Technology. His research interests include the optimal integration of the power and gas systems, and the storage across multiple energy carriers.

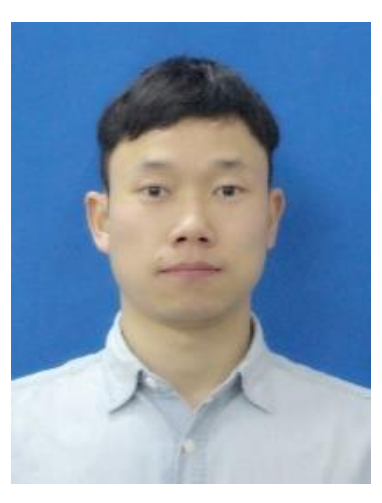

**Shichang Cui** (Member, IEEE) received the B.S. degree in Automation and the Ph.D. degree in control science and engineering from the Huazhong University of Science and Technology (HUST), Wuhan, China, in 2016 and 2021, respectively.

He currently works as a Postdoctoral Fellow in the State Key Laboratory of Advanced Electromagnetic Engineering and Technology, School of Electrical and Electronic Engineering, Huazhong University of Science and Technology, Wuhan 430074, China. From Sep. 2019 to Sep. 2020, he was visiting the department of Mechanical Engineering, University of Victoria, Canada, supported by the CSC Joint Doctoral Program. His current research interests include stochastic optimization, distributed optimization, game theory, and energy management for smart grids.

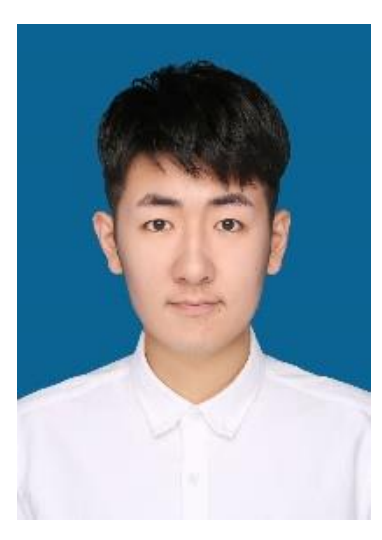

**Shengshi Wang** (Student Member, IEEE) received the B.Eng. degree in electrical engineering from the Chongqing University, Chongqing, China, in 2020. He is currently pursuing the Ph.D. degree in electrical engineering with the Huazhong University of Science and Technology, Wuhan, China.

His research interests include the optimal operation of the integrated energy transmission network.

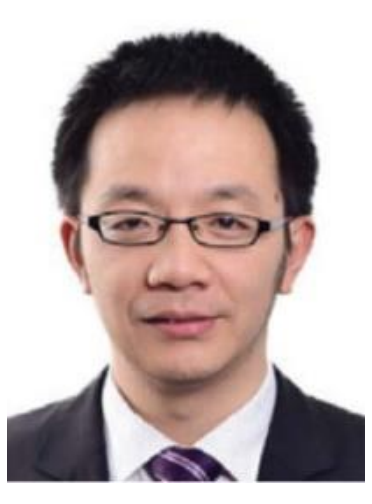

**Wei Yao** (Senior Member, IEEE) received the B.S. and Ph.D. degrees in electrical engineering from the Huazhong University of Science and Technology (HUST), Wuhan, China, in 2004 and 2010, respectively.

He was a Postdoctoral Researcher with the Department of Power Engineering, HUST, from 2010 to 2012 and a Postdoctoral Research Associate with the Department of Electrical Engineering and Electronics, University of Liverpool, Liverpool, U.K., from 2012 to 2014. He is currently a Professor with the School of Electrical and Electronics Engineering, HUST. His current research interests include power system stability analysis and control, renewable energy, HVDC and DC Grid, and application of artificial intelligence in smart grid.

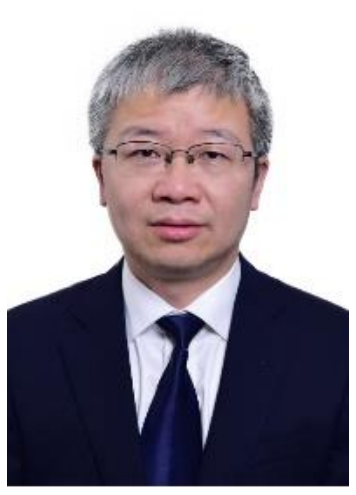

**Jinyu Wen** (Member, IEEE) received the B.S. and Ph.D. degrees in electrical engineering from the Huazhong University of Science and Technology, Wuhan, China, in 1992 and 1998, respectively.

He was a Visiting Student from 1996 to 1997, and a Research Fellow from 2002 to 2003, with the University of Liverpool, Liverpool, U.K., and a Senior Visiting Researcher with the University of Texas at Arlington, Arlington, TX, USA, in 2010. From 1998 to 2002, he was a Director Engineer with XJ Electric Company Ltd., China. In 2003, he joined HUST, where he is currently a Professor with the School of Electrical and Electronics Engineering. His current research interests include renewable energy integration, energy storage, multiterminal HVDC, and power system operation and control.